\documentclass[twocolumn]{article}

\usepackage{graphicx}
\usepackage{amsmath}
\usepackage{bbold}
\usepackage{braket}

\usepackage{authblk}
\usepackage{hyperref}

\begin{document}

    \title{Tutorial: Cavity Quantum Optomechanics}
    \author{Amarendra K. Sarma\thanks{aksarma@iitg.ac.in} ~and Sampreet Kalita\thanks{sampreet@iitg.ac.in}}
    \affil{\textit{Department of Physics, Indian Institute of Technology Guwahati, Guwahati-781039, Assam, India}}
    \date{}
    
    \maketitle

    \begin{abstract}
    	Exploring quantum physics in macroscopic systems and manipulating these systems for various technological applications has been a topic of intense research in the last one decade or so.
        In this regard, the field of cavity quantum optomechanics turns out to be one of the most rapidly emerging area of research.
        It has opened many doors to study various open ended fundamental questions in quantum physics, apart from numerous possible applications.
        A typical cavity optomechanical system consists of two mirrors, one fixed while the other one is movable.
        These systems may be of micrometer or nano-meter in dimensions.
        The electromagnetic radiation incident on the system may get coupled to the mechanical motion of the movable mirror.
        This opto-mechanical coupling is the root of all phenomena such as quantum entanglement, state-transfer, squeezing and so on.
        In this short tutorial, basic concepts of cavity quantum optomechanics are discussed.
        We hope that this tutorial would motivate readers, both theorists and experimentalists, to take up advanced studies in this immensely fruitful area of research. 
    \end{abstract}

    \tableofcontents


    \section{Introduction}
        \label{sec:intro}
        Cavity quantum optomechanics is a rapidly evolving research area of physics, and science in general \cite{RevModPhys.86.1391, SBH.CavityOptomechanics.Aspelmeyer}.
        This field of research requires utilization of the tools of quantum optics in a variety of condensed matter systems.
        In the rest of the article, we use the term `optomechanics' frequently, to mean cavity optomechanics or cavity quantum optomechanics.
        Optomechanics deals how light couples with mechanical motion.
        There are mechanical systems, such as beams, cantilever etc, on micro and nano scale, that can vibrate in the frequency range of kHz to GHz.
        Over the past two decades or so, researchers have learnt how to control these systems, i.e. to couple them and read out their motion, by a variety of methods.
        One of the first such approach was to couple these motions to electrical circuits, resulting in the so-called nano electrical mechanical systems (NEMS) \cite{RevSciInstrum.76.061101}.
        Later, similar endeavours led to couple these systems to electromagnetic field in the optical domain, giving birth to the area of optomechanics.
        At the root of optomechanical interaction is the so-called radiation pressure induced interaction between photons and mechanical motion in a cavity \cite{PhysikZ.10.817, SovPhysJETP.25.653}.
        The radiation pressure force, arising due to the momentum carried by light, can displace a movable end mirror of the cavity.
        This in turn changes the length of the cavity, resulting in modification of the cavity frequency.
        With the rapid advances in micro and nanofabrication techniques, this nonlinear interaction has led to the exploration of a wide variety of interesting phenomena both theoretically and experimentally, such as squeezing of the light field and mechanical motion, entanglement between optical and mechanical modes, bistability, optomechanical normal mode splitting, optomechanically induced transparency, and so on.
        
        Optomechanics turns out to be potential tool to probe fundamental physics and to exploiting its various concepts for various applications, primarily in the so-called quantum technologies \cite{NatPhys.18.15}.
        Optomechanics can be used to test quantum mechanics in an entirely new domain.
        It is possible to produce non-classical states of heavy mechanical objects and test Quantum Mechanics.
        One of the fascinating topics of research where optomechanics could be useful is to get an idea about so-called transition boundary between the  classical world to quantum world or vice versa.
        To put in a simple language, to know, when to stop using Newton's laws and use Schrodinger's formalism etc.
        For heavy objects we generally do not talk about quantum mechanics.
        As we go to larger and larger objects they couple to the unavoidable fluctuations of noisy environment more strongly resulting in the destruction of superposition and leading to decoherence.
        If the objects are large enough decoherence may not even allow us to produce superposition of states in the first place.
        One primary goal is to test how \textit{decoherence} evolves for heavy objects.
        There is a speculation, primarily due to British mathematician and physicist Roger Penrose, that if we go to heavier objects additional mechanism for decoherence might kick in owing to gravitation!
        This may enable us to look into various aspects of quantum gravity \cite{TaylorFrancis.QuantumOptomechanics.Bowen}!

        The reason for enormous popularity of optomechanical systems (OMS) and its variants is numerous possible applications, primarily in quantum technologies \cite{NatPhys.18.15, AVSQuantumSci.3.015901}.
        OMS may be used to for quantum information processing, i.e. to store quantum information and transfer it.
        One can couple a super-conducting qubit to a mechanical system and then couple it to an optical system to process the information.
        It is important to note that to study the quantum states of such systems, we need to cool these mechanical systems.
        Because their vibration frequency $\omega_{m}$ lies in the range of kHz-GHz, which corresponds to temperature $T \approx \hbar \omega_{m} / k_{B}$ that lies far below $20$ mK or so, we need to look for some clever methods to cool mechanical systems.
        It is good to reduce thermal fluctuation but not always necessary to go to quantum ground state as regards applications are concerned.
        As an example, for the last 20 years or so, cantilevers are used in atomic force microscopy.
        The idea is to use these cantilevers to scan across the surface of some material and look at the surface of the material with atomic resolution.
        Ultra sensitive detection of tiny forces (force gradients to be precise.
        Because if we apply a force gradient to an oscillator it adds some effect to the spring constant of the oscillator and shifts the frequency which is easy to detect), masses, displacement etc. is also demonstrated using OMS.
        It is well known that interaction of light field and the mechanical system automatically provide some nonlinearity.
        So, whenever we have a dynamical system with nonlinearity, by proper driving we can turn it into an amplifier.
        Thus, we can amplify very small signal, and indeed OMS are used in signal amplification and processing.
        \textit{The best thing about optomechanical systems is that everything could be integrated on a chip} and nano-fabrication of these devices is possible \cite{PhysicsToday.65.29}.
        The primary advantages of optomechanical platforms over that of other ones are due to its small size, high quality factor (as a result, information can be stored for a very long time) and its integrability with various other systems \cite{NatNanotechnol.13.11}.
        Moreover, mechanical system can be coupled to virtually anything \textemdash{} qubits, superconducting circuits, spins, cold atoms etc.
        
        It is obvious that, owing to the enormous significance of the field of cavity optomechanics, and the fact that the field is developing pretty fast since the last decade, there are numerous review articles, short notes, book chapters and so on.
        Even, there are quite a few online materials by experts.
        The objective of this tutorial is to give the readers a comprehensive introduction to the key concepts of cavity quantum optomechanics, which would surely motivate them to take up further studies.


    \section{Mechanical Effects of Light}
        \label{sec:effects}
            The radiation pressure force is at the root of all optomechanical phenomena.
            In fact, radiation pressure force is behind the most well known optical tools called optical tweezers, frequently used by biologists.
            Arthur Ashkin of Bell Labs contributed significantly towards understanding of radiation pressure force \cite{PhysRevLett.40.729}. 
            \begin{figure}[h!]
                \centering
                \includegraphics[width=0.48\textwidth]{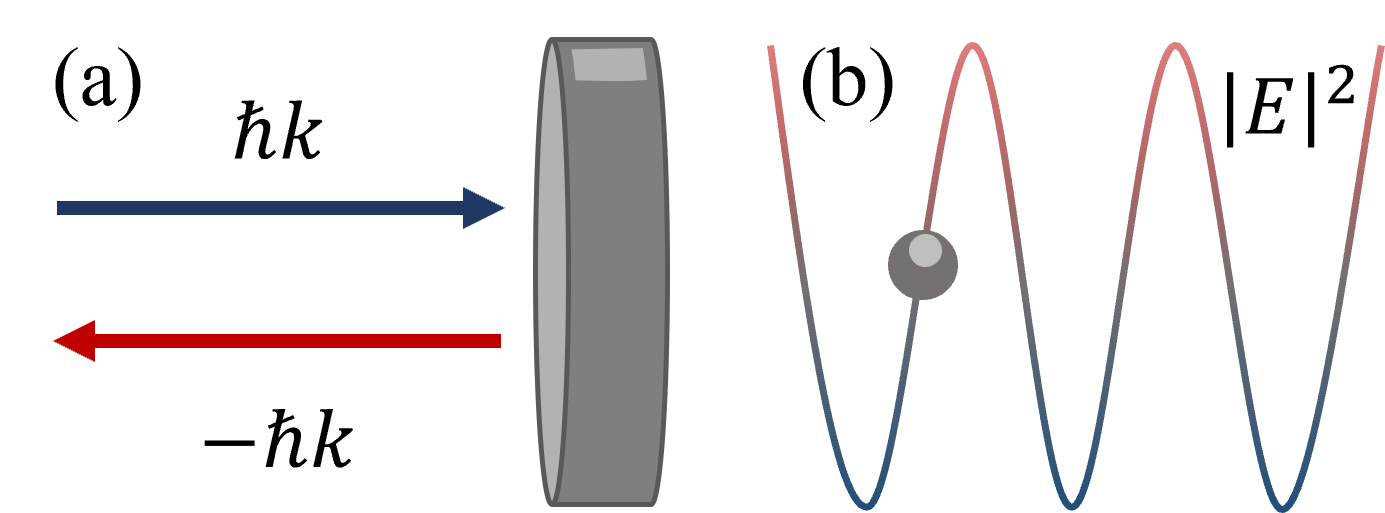}
                \caption{Two examples of mechanical effects of light: (a) radiation pressure force and (b) dipole force.}
                \label{fig:effects}
            \end{figure}

        \subsection{Radiation Pressure Force}
            \label{sec:effects_rpf}
            A typical set-up to explain the radiation pressure force is depicted in Figure \ref{fig:effects} (a).
            A photon incident on a perfectly reflecting mirror gives a momentum kick of $\hbar k$ to the mirror and while it gets reflected experiences a momentum kick, $- \hbar k$ from the mirror.
            Thus it experiences an overall momentum change given by the equation
            \begin{equation}
                \label{eqn:effects_momentum}
                \Delta p = 2 \hbar k = 2 \frac{\mathcal{E}}{c}.
            \end{equation}
            Here $\mathcal{E}$ is the energy of the single photon.
            If we have a steady stream of photons then the force will be:
            \begin{equation}
                \label{eqn:effects_rpf}
                F_{RP} = \frac{N_{p}}{t} \Delta p = 2 \frac{P}{c}.
            \end{equation}
            Here $N_{p}$ is the number of photons impinging on the mirror and $P$ refers to power.
            For example, the radiation pressure force due to sunlight is nearly on the order of $10^{-5}$ Newtons on one square meter, a very tiny force!
        
        \subsection{Dipole Force}
            \label{sec:effects_dipole}
            Dipole force is another example of a radiation pressure force.
            Take two counter propagating light waves.
            They will form a standing wave, where the intensity of the resultant wave will vary sinusoidally.
            If an atom or any polarizable object is placed in the light wave, then the electric field will induce some dipole moment in the atom.
            The atom, in turn then will interact with the electric field, $\vec{E} (t)$.
            The energy, $\mathcal{E}$, of the dipole placed in the electric field is given by:
            \begin{equation}
                \label{eqn:effects_dipole}
                \mathcal{E} = - \vec{\mu} (t) \cdot \vec{E} (t) \propto \pm \left| E \right|^{2},
            \end{equation}
            the dipole moment being proportional to the electric field intensity.
            The sign depends on the direction of the laser electric field and the dipole moment.
            That again depends on how laser frequency lies with respect to the atomic transition frequency.
            When the sign is negative (laser light is red-detuned) the atom will be dragged towards higher intensity.
            On the other hand, if the laser light is blue-detuned; atom will be pushed towards lower intensity.
            This technique is used to trap atoms or molecules or polarizable particles: Atoms in standing light wave is called \textit{Optical lattice}.


    \section[Generic Model of an Optomechanical System]{Generic Model of an \\Optomechanical System}
        \label{sec:oms}
        \begin{figure*}[!ht]
            \centering
            \includegraphics[width=0.84\textwidth]{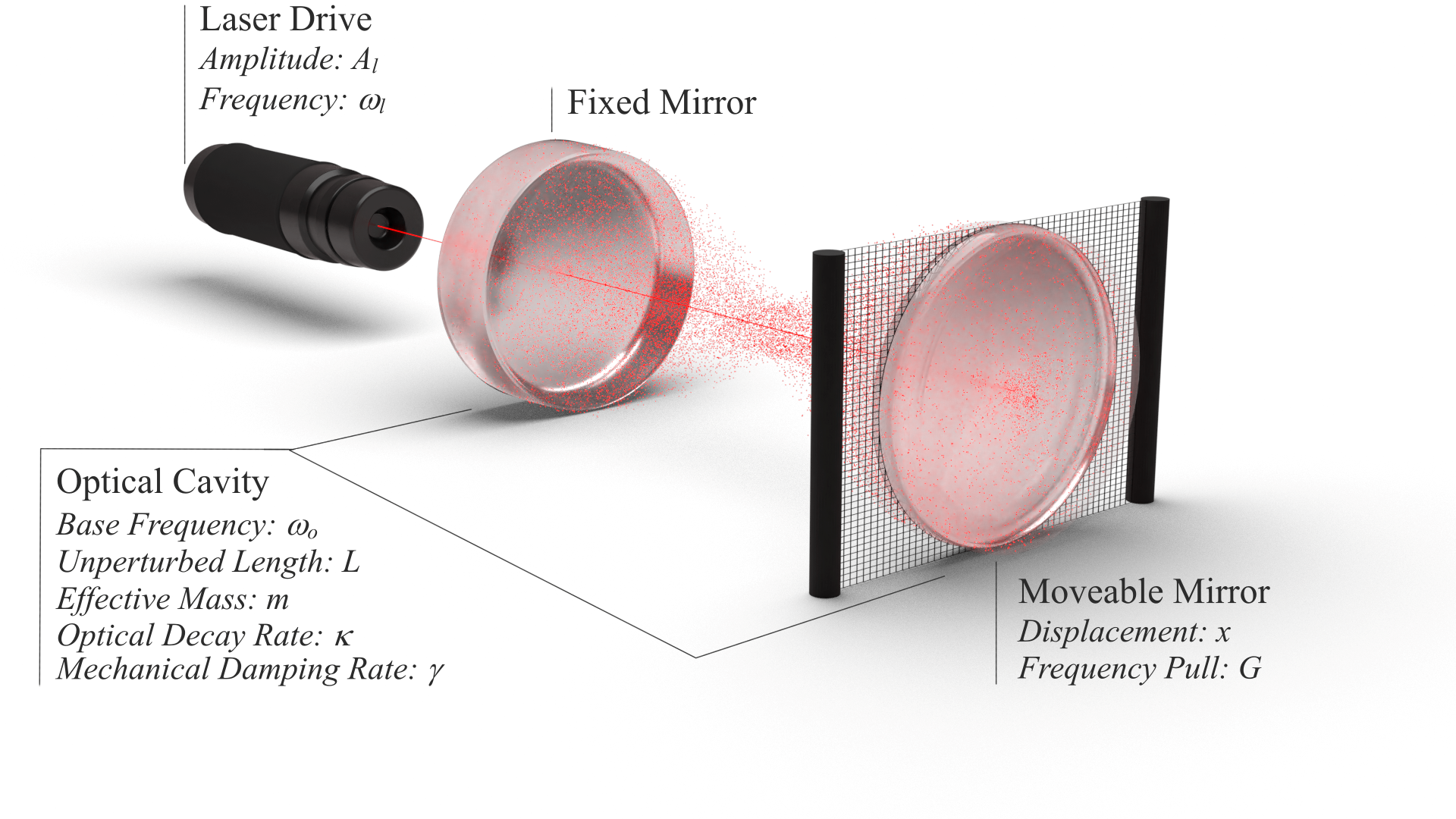}
            \caption{
                Illustration of a typical optomechanical cavity with relevant parameters.
                Source: \cite{AVSQuantumSci.3.015901}.
            }
            \label{fig:oms_model}
        \end{figure*}

        Although there are numerous platforms where optomechanical effects could be realized, a surprisingly simple model can be used to describe all the phenomena displayed by those seemingly different platforms.
        Such a model, as depicted in Figure \ref{fig:oms_model}, is a cavity with two mirrors \textemdash{} one fixed and the other moveable, separated by some distance (say $L$) \textemdash{} and driven by a laser source (say, with amplitude $A_{l}$ and frequency $\omega_{l}$) \cite{RevModPhys.86.1391}.
        The cavity boosts the light field injected inside it by the laser and the radiation pressure force of the intracavity photons results in a continuous transfer of momentum to the mechanically-compliant end-mirror.
        This gives rise to mechanical degrees of freedom, whose individual movements are given by the response of elastic strain in their harmonic modes \cite{TaylorFrancis.QuantumOptomechanics.Bowen}.
        But the macroscopic displacement of the mirror largely depends on the shorter-wavelength microscopic phononic modes, and the collective degrees of freedom can be described by a single mechanical mode.
        Hence, the mechanical mirror can be approximated to oscillate with an effective resonance frequency of $\omega_{m}$ and an effective mass of $m$.
        One can then focus on that single optical mode of the cavity (say with frequency $\omega_{o} = n \pi c / L$ with mode number $n$) which is close to the laser frequency.
        For a mechanical displacement of $\hat{x}$, the new cavity resonance frequency can be expressed as
        \begin{equation}
            \label{eqn:oms_omega}
            \omega_{c} (\hat{x}) = \frac{n \pi c}{L + \hat{x}} = \frac{\omega_{o}}{1 + \hat{x} / L} \approx \omega_{o} \left( 1 - \frac{\hat{x}}{L} \right).
        \end{equation}

        This relationship gives us the \textit{frequency pull} parameter \cite{RevModPhys.86.1391} $G = \partial \omega_{c} (\hat{x}) / \partial \hat{x} = \omega_{o}/ L$, which quantifies the linear dispersive shift of the resonance frequency of the optical field induced by the mechanics.
        It is important to note here that a driven cavity with two fixed mirrors only accommodates modes having frequencies in integral multiples of $\pi c / L$ \textemdash{} a parameter called the \textit{free spectral range} (FSR) of the cavity \cite{SBH.CavityOptomechanics.Aspelmeyer}.
        For most optomechanical systems, this value is much larger than the mechanical frequency.

        \subsection{Hamiltonian Formalism}
            \label{sec:oms_h}
            The total Hamiltonian, $H$, of a typical optomechanical system consists of the contributions from the optical components, the mechanical components, the interaction between the optics and the mechanics, the laser drive and the transfer of photons and phonons to and from the environment.
            Let us derive each of these terms one by one.

            The optical and mechanical modes can be represented as harmonic oscillators with annihilation operators $\hat{a}$ and $\hat{b}$ respectively.
            Each of these operators obey the Bosonic commutation relations $[ \hat{\mathcal{O}}, \hat{\mathcal{O}}^{\dagger} ] = 1$ with their number operators given by $\hat{n}_{\mathcal{O}} = \hat{\mathcal{O}}^{\dagger} \hat{\mathcal{O}}$.
            Thus, one can write the energy of the optical (mechanical) mode in terms of the number of photons (phonons) times the energy of each quanta, and obtain the combined Hamiltonian of the optical and mechanical modes as
            \begin{subequations}
                \label{eqn:oms_h_om}
                \begin{eqnarray}
                    \hat{H}_{om} & = & \hbar \omega_{c} (\hat{x}) \hat{a}^{\dagger} \hat{a} + \hbar \omega_{m} \hat{b}^{\dagger} \hat{b}, \\
                    & = & \hbar \omega_{o} \hat{a}^{\dagger} \hat{a} + \hbar \omega_{m} \hat{b}^{\dagger} \hat{b} + \hat{H}_{int},
                \end{eqnarray}
            \end{subequations}
            where we have used Equation \eqref{eqn:oms_omega} to write the Hamiltonian describing the interaction between the optical and mechanical modes as
            \begin{equation}
                \label{eqn:oms_h_int}
                \hat{H}_{int} = - \hbar G \hat{x} \hat{a}^{\dagger} \hat{a} = - \hbar g_{0} \left( \hat{b}^{\dagger} + \hat{b} \right) \hat{a}^{\dagger} \hat{a}.
            \end{equation}

            Here, we have introduced the \textit{optomechanical coupling constant} \cite{RevModPhys.86.1391}
            \begin{equation}
                \label{eqn:oms_h_g0}
                g_{0} = G x_{ZP} = G \sqrt{\frac{\hbar}{2 m \omega_{m}}},
            \end{equation} 
            where $x_{ZP}$ is the zero-point position and is typically of the order of $10^{-15}$ m.
            This coefficient relates the radiation pressure force of a single photon on the position of a phonon and can range from a few Hz to a few MHz.
            \\

            \textbf{Exercise 1:} Obtain the radiation pressure force and its relationship with the frequency pull parameter using the Hamiltonian in Equation \eqref{eqn:oms_h_om}.
            \\

            We now add the laser drive term which describes a photon created inside the cavity at the laser frequency and its Hermitian conjugate.
            Thus, we obtain the effective Hamiltonian for the system as
            \begin{eqnarray}
                \label{eqn:oms_h}
                \hat{H}_{sys} & = & \hbar \omega_{o} \hat{a}^{\dagger} \hat{a} + \hbar \omega_{m} \hat{b}^{\dagger} \hat{b} - \hbar g_{0} \hat{a}^\dagger \hat{a} \left( \hat{b}^{\dagger} + \hat{b} \right) \nonumber \\
                && + i \hbar A_{l} \left( \hat{a}^{\dagger} e^{- i \omega_{l} t} - \hat{a} e^{i \omega_{l} t} \right).
            \end{eqnarray}

            The time-dependent terms in this Hamiltonian can be removed by switching to the rotating frame of the laser.
            This transformation also helps us to analyze the slower dynamics of the mechanical motion.
            We obtain the Hamiltonian in this frame as
            \begin{eqnarray}
                \label{eqn:oms_h_l}
                \hat{H}_{sys} & = & - \hbar \Delta_{0} \hat{a}^{\dagger} \hat{a} + \hbar \omega_{m} \hat{b}^{\dagger} \hat{b} - \hbar g_{0} \hat{a}^{\dagger} \hat{a} \left( \hat{b}^{\dagger} + \hat{b} \right) \nonumber \\
                && + i \hbar A_{l} \left( \hat{a}^\dagger - \hat{a} \right),
            \end{eqnarray}
            where $\Delta_{0} = \omega_{l} - \omega_{o}$ is the \textit{laser detuning} and the operator $\hat{a}$ is written in the rotating frame.
            \\

            \textbf{Exercise 2:} Choose an appropriate unitary transformation and derive Equation \eqref{eqn:oms_h_l}.
            \\

            Typically, the resonant frequency of the cavity ($\omega_{o}$) is on the order of $10^{15}$ Hz, whereas the mechanical resonance frequency ($\omega_{m}$) is usually in the range of MHz to GHz.
            The laser frequency is therefore adjusted in such a way that its detuning $\Delta_{0}$ is comparable to $\omega_{m}$.
            Also, since the optical frequencies are very high, the effect of the thermal photons entering the cavity can be safely neglected and the cavity can be considered to be coupled with a reservoir at zero temperature.
            However, the laser drive does introduce noise into the system.
            On the other hand, thermal phonons play a very important role on the quantum dynamics of the mechanical mode.
            We explore these environmental effects systematically in the next section.
            Also, note that  we shall drop the operator symbol in the rest of the article to make the expressions simpler.

        \subsection{Quantum Langevin Equations}
            \label{sec:oms_qle}
            Optomechanical cavities are open quantum systems.
            Such systems interact with the environment and undergo fluctuations and dissipations \cite{TaylorFrancis.QuantumOptomechanics.Bowen, PhysRevA.31.3761}, resulting in the decay of photons, mechanical damping as well as insertion of environmental noises.

            \subsubsection{Dissipation Processes}
                \label{sec:oms_qle_diss}
                Since the photons are coupled to the external environment through the mirrors, they experience absorption and scattering losses.
                These losses collectively define the \textit{optical decay rate} $\kappa$.
                The optical decay rate also determines the quality factor of the cavity, which is given by $Q_{o} = \omega_{o} / \kappa$.
                This factor signifies the total number of times a single photon oscillates inside the cavity before moving out.
                Similarly, the mechanical motion is affected by environmental factors such as viscous drag, clamping losses, phonon-phonon interactions and losses due to its composition.
                The \textit{mechanical damping rate} $\gamma$ takes these effects into account.
                It also characterizes the strength of coupling between the mechanical mode and the environment.
                The mechanical quality factor $Q_{m} = \omega_{m} / \gamma$ is equivalently related to the phonon lifetime.

            \subsubsection{Input Fluctuations}
                \label{sec:oms_qle_fluct}
                As mentioned earlier, the environment also introduces noises into the system and steers the system dynamics.
                However, such a thermal environment could be modelled by a bath of several non-interacting harmonic oscillators.
                Effectively, the noises entering the system can also be approximated as \textit{Markovian}, zero-mean and $\delta$-correlated.
                It is useful to note that a system is termed as Markovian if its present state does not depend on previous history.
                Moreover, zero-mean and $\delta$-correlated noise are known as \textit{white noise}.
                Therefore, the white noise entering the cavity through the laser drive ($a_{in}$) and the \textit{Langevin noise} introduced into the mechanical oscillations ($b_{in}$) follow the standard correlation relations \cite{TaylorFrancis.QuantumOptomechanics.Bowen}:
                \begin{subequations}
                    \label{eqn:oms_qle_bcr}
                    \begin{eqnarray}
                        \langle a_{in} (t) a_{in}^{\dagger} (t') \rangle & = & \delta (t - t'), \\
                        \langle b_{in}^{\dagger} (t) b_{in} (t') \rangle & = & n_{th} \delta (t - t'), \\
                        \langle b_{in} (t) b_{in}^{\dagger} (t') \rangle & = & \left( n_{th} + 1 \right) \delta (t - t'),
                    \end{eqnarray}
                \end{subequations}
                where $n_{th} = [ \mathrm{exp} \{ \hbar \omega_{m} / ( k_{B} T ) \} - 1]^{-1}$ is the mean thermal occupancy at bath frequency $\omega_{m}$ and temperature $T$.
                $k_{B}$ is the Boltzmann constant. 

            \subsubsection{Input-Output Relations}
                \label{sec:oms_qle_ior}
                Although the photon \textit{energy} decay rate $\kappa$ is a collective term, it is predominantly associated with the input-output losses than the intrinsic cavity losses.
                As such, the optical field \textit{amplitude} effectively decays at the rate $\kappa/2$.
                Also, it is important to note here that input-output theory relates the laser power $P_{l}$ entering the cavity to the amplitude inside it as $A_{l} = \sqrt{\kappa P_{l} / \hbar \omega_{l}}$.
                The input power associated with the white noise is given by $P_{in} = \hbar \omega_{l} \langle a_{in}^{\dagger} a_{in} \rangle$.
                With the mean-field approximation for this input noise, it follows that the optical mode is modulated by an extra amount of $\sqrt{\kappa} a_{in}$.
                The input-output relation is then obtained as \cite{PhysRevA.31.3761}:
                \begin{equation}
                    \label{eqn:com_qle_ior}
                    a_{out} = a_{in} - \sqrt{\kappa} a,
                \end{equation}
                where $a_{out}$ is the measured field exiting the cavity through the mirror.

            \subsubsection{Equations of Motion}
                \label{sec:oms_qle_eom}
                Using the Hamiltonian of the system derived in Equation \eqref{eqn:oms_h_l} and the expressions for fluctuations and dissipations, we obtain the quantum Langevin equations (QLEs) corresponding to the cavity field and the mechanical mode as \cite{TaylorFrancis.QuantumOptomechanics.Bowen}:
                \begin{subequations}
                    \label{eqn:oms_qle_eom}
                    \begin{eqnarray}
                        \dot{a} & = & \frac{1}{i \hbar} \left[ a, H_{sys} \right] - \frac{\kappa}{2} a + \sqrt{\kappa} a_{in} \\
                        & = & - \frac{\kappa}{2} a + i \Delta_{0} a + i g_{0} \left( b^{\dagger} + b \right) a \nonumber \\
                        && + A_{l} + \sqrt{\kappa} a_{in}, \\
                        \dot{b} & = & \frac{1}{i \hbar} \left[ b, H_{sys} \right] - \frac{\gamma}{2} b + \sqrt{\gamma} b_{in} \\
                        & = & - \frac{\gamma}{2} b + i \omega_{m} b + i g_{0} a^{\dagger} a + \sqrt{\gamma} b_{in}.
                    \end{eqnarray}
                \end{subequations}
                \\

                \textbf{Exercise 3:} Using the Bosonic commutation relations, derive Equations \eqref{eqn:oms_qle_eom}.
                \\

                It can be seen here that the frequency shift introduced in the optical mode by the mechanical position changes its amplitude, which in turn changes the mechanical mode amplitude.
                Both of these phenomena depend directly on the optomechanical coupling constant $g_{0}$.
                Along with this, the losses and noises introduce a feedback-like behaviour into the system known as the \textit{dynamical optomechanical backaction} \cite{SBH.CavityOptomechanics.Aspelmeyer}.
                We shall revisit these coupled equations and their associated phenomena in later sections.
                But before getting into the intricacies, let us briefly discuss the qualitative aspects of the optomechanical interaction.

    
    \section[Basic Physics of Optomechanical Systems]{Basic Physics of \\Optomechanical Systems}
        \label{sec:basic}        
        \subsection{Statics}
            \label{sec:basic_statics}
\begin{figure}[h!]
                \centering
                \includegraphics[width=0.48\textwidth]{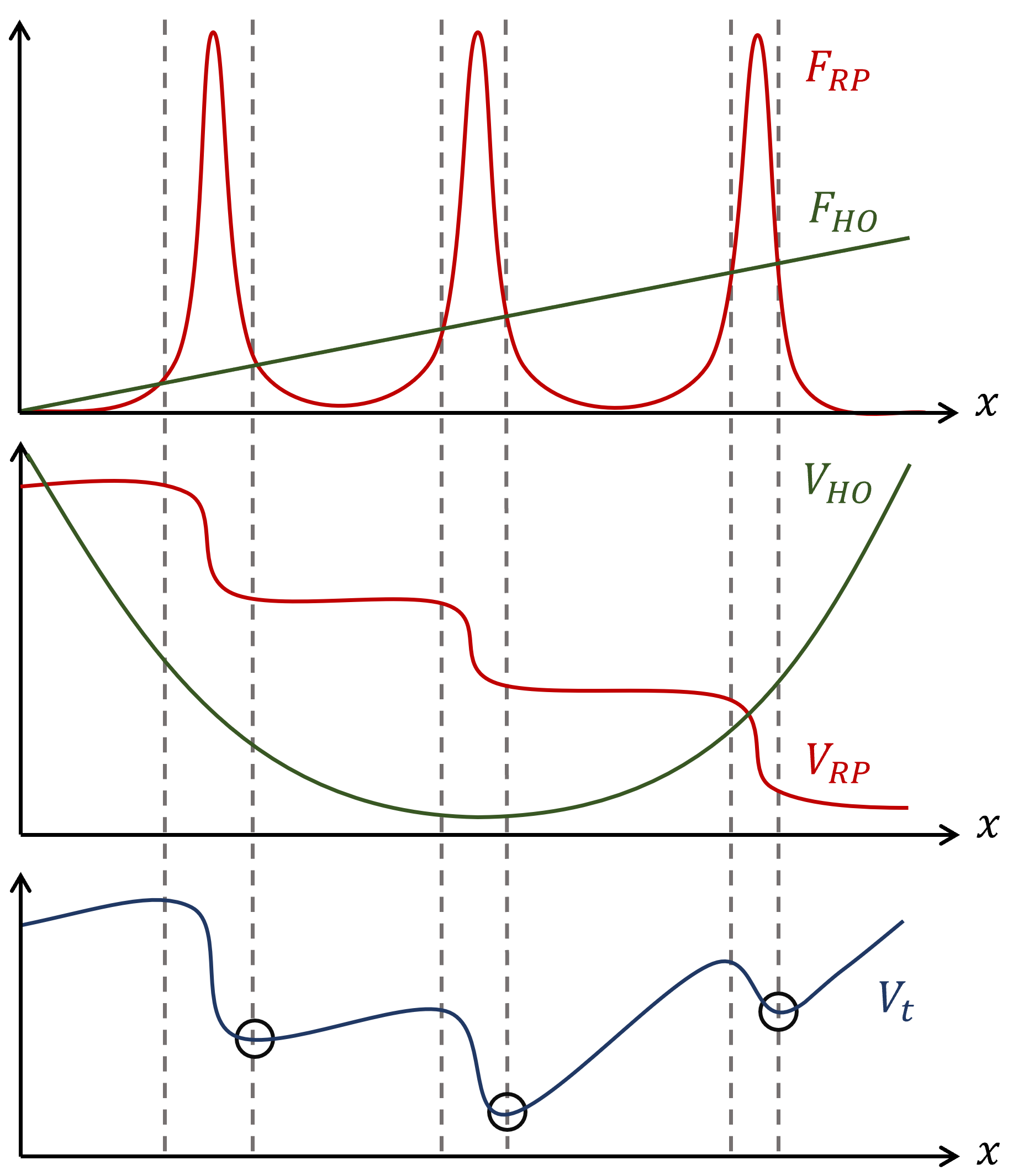}
                \caption{
                    \textit{Top:} Radiation pressure force ($F_{RP}$) and the intrinsic restoring force ($F_{HO}$) versus displacement ($x$).
                    \textit{Center:} Radiation pressure potential ($V_{RP}$) and intrinsic harmonic potential ($V_{HO}$) versus $x$.
                    \textit{Bottom:} Total potential ($V_{t}$) versus $x$.
                    The circles denote stable equilibrium positions.
                }
                \label{fig:basic_statics}
            \end{figure}

            Let us assume that the mechanical mirror is moved very, very slowly, akin to a static situation, i.e. for each position of the movable mirror, we give enough time to the mechanical oscillator to settle to the new state and in particular, for the light field intensity to adjust to the new length of the cavity.
            Now, the question is, what happens to the light field?
            As the mirror moves along, at some point it will be in resonance with the incoming laser light and the cavity will be filled up with light.
            So the radiation pressure force will also increase.
            We can plot the radiation pressure force (or equivalently the circulating light intensity) versus the displacement, as depicted in Figure \ref{fig:basic_statics}.
            One can see a typical Fabry-Perot Lorentzian resonance line shape. The spacing between two such resonances is given by $\lambda / 2$. The full width at the half-maximum (FWHM) of the resonance is given by, $\frac{\lambda}{2 \mathcal{F}}$, where ${\mathcal{F}}$ is the so-called Finesse of the cavity.
            The corresponding potential versus displacement could also be plotted.

            It is clear that when there is no force, the potential vs. displacement curve has to be horizontal.
            When there is a force, the force pushes the mirror to the right, so we will have a dip and so on.
            This is a rather curious potential induced by the radiation pressure force.
            Please note that the full mechanical motion sees the standard restoring force potential plus this potential.
            One can see that at some places there is local minima.
            There are several such minima depending on the strength of the extra potential, provided by the light field intensity.
            The local minima of the overall potential as seen by the mechanical system correspond to stable equilibrium positions.
            The system can sit in any of these local minima depending on its history.

            By changing some parameters, say the light intensity and looking at the output intensity, one can see that the system oscillates around a local minimum.
            However, it may also enter into a different local minimum and now if we reduce the light intensity it may stay in that minima rather than going back to the original minimum, giving rise to the phenomena of Hysteresis.
            This is happening owing to the existence of many equilibrium positions.
            Note that the original curvature of the potential (without the light field) gets changed owing to the presence of the light field.
            So the spring constant of the mechanical oscillator now has an effective spring constant, as expressed in Equation \eqref{eqn:basic_v}.
            This is known as the optical spring effect.
            \begin{equation}
            	\label{eqn:basic_v}
                V_{t} = V_{HO} + V_{RP} = \frac{1}{2} K_{eff} x^{2}.
            \end{equation}

        \subsection{Dynamics}
            \label{sec:basic_dynamics}
\begin{figure}[h!]
                \centering
                \includegraphics[width=0.48\textwidth]{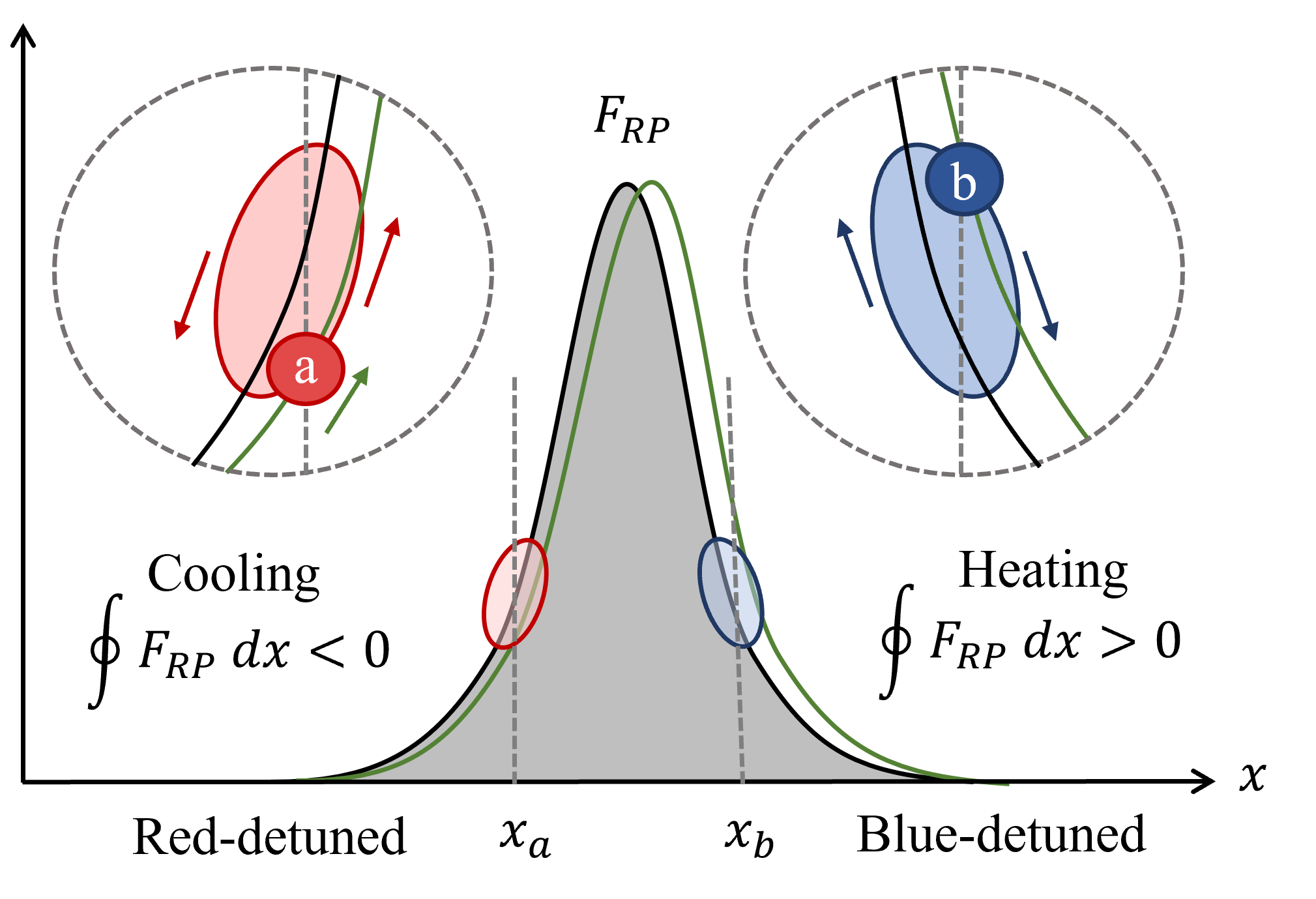}
                \caption{
                    Radiation pressure force ($F_{RP}$) versus displacement ($x$) depicting cooling and heating for red-detuned and blue-detuned laser respectively.
                    The green line denotes the effect of time-lag.
                }
                \label{fig:basic_dynamics}
            \end{figure}

            Let us now relax the static case.
            The mechanical oscillator or the cantilever now moves with a finite speed.
            The light field does not get time to completely track the mechanical motion.
            So we will have time-lag effects.
            Let us imagine the following situation.
            If we move the mirror adiabatically, i.e. with zero speed, then we obtain the usual Lorentzian profile.
            Now, say we sweep along with a finite speed (refer to the solid green curve in Figure \ref{fig:basic_dynamics}).
            Then, as we reach the position $x_{a}$, the intensity has not built up yet to its full value, so we observe a smaller intensity, a smaller force.
            Again, as we reach position $x_{b}$, the intensity has not yet relaxed down to small value but retains memory of having been larger intensity before.
            So we will observe a displaced profile.
            So there will be a time lag between the motion and the observed intensity.
            The time scale of this time-lag is set by the cavity ring down rate, given by $\kappa$, which is also the photon decay rate.
            
            In the dynamic case, the force is no longer a function of position at a given instant of time as was the case with static:
            \begin{equation}
                F_{RP} \neq F_{RP} (x (t)).
            \end{equation}

            But $F_{RP} (t)$ , the force at a given time depends on the full pre-history of the motion of x because that determines how the light field builds up and depletes again.
            \begin{equation}
                F_{RP} = F_{RP} (x (t^{\prime}), t^{\prime} \leq t).
            \end{equation}

            This force is no longer a conservative force.
            So we cannot write down a potential.
            It implies this force generates some friction.
            What are the consequences then?

           To understand this, let us revert back to the resonance curve for the adiabatic case, denoted by the solid black curve in Figure \ref{fig:basic_dynamics}.
            It may be noted here that even in thermal equilibrium, the harmonic oscillator undergoes some oscillations whose amplitudes change with time.
            Let us follow through one such cycle of these oscillations, say at position $x_{a}$.
            As we approach the resonance, the force is little bit smaller than earlier due to time lag (as it still had to build up) and then as we go back, we have already reached the higher intensity and the force will remain high for some time than the adiabatic case.
            So the force in the two halves will not be identical in magnitude.
            So as a consequence, if  we calculate the work done by the radiation pressure force in such a cycle, it will turn out to be non-zero.
            In fact,
            \begin{equation}
                \oint F_{RP} dx < 0.
            \end{equation}

            This means, the force provided by the light field extracts energy from the mechanical motion, inducing extra damping in the mechanical oscillator.
            This is called light induced damping or optomechanical damping, $\gamma_{om}$.
            So, the total damping of the mechanical oscillator will be due to both the mechanical damping ($\gamma$) and the optomechanical damping ($\gamma_{om}$).
            \begin{equation}
                \gamma_{t} = \gamma + \gamma_{om}.
            \end{equation}
            
            Now, whenever there is extra damping but no extra fluctuations, then this damping can simply be used to damp away the thermal fluctuations.
            Thermal fluctuations always arise as an equilibrium between damping that wants to extract energy and thermal random force that comes from outside that tries to heat up the system.
            In thermal equilibrium we have a balance between the two.
            So if we increase the damping we will be able to effectively reduce the thermal fluctuation. 
            \begin{eqnarray}
                \oint F_{RP} dx < 0 \Rightarrow \text{extra damping} \nonumber \\
                \Rightarrow \text{light-induced cooling}.
            \end{eqnarray}
            
            This is good news!
            Because the typical refrigerators in Lab are not good enough to cool mechanical systems down to the ground state.
            We can then exploit these extra mechanisms to cool the oscillator further down.
            Note that, the situation we have considered above has a cavity length smaller compared to the resonance one.
            This implies that: $\omega_{o} > \omega_{l}$, i.e. the laser is red-detuned.
            If we place the cantilever in the position, $x_{b}$ (please refer to Figure \ref{fig:basic_dynamics}), we will have a situation where the cavity length is longer than the resonance one.
            It will correspond to the case, $\Delta_{0} > 0$, i.e. the laser is blue detuned.
            In this configuration, the laser light will dump energy to the mechanical mirror, thereby heating it.
            Thus,
            \begin{subequations}
                \begin{eqnarray}
                    \label{eqn:basic_dynamics}
                    && \Delta_{0} < 0: \text{Cooling}; \quad \gamma_{om} > 0, \\
                    && \Delta_{0} > 0: \text{Heating}; \quad \gamma_{om} < 0.
                \end{eqnarray}
            \end{subequations}


    \section[Classical Regime of Cavity Optomechanics]{Classical Regime of \\Cavity Optomechanics}
        \label{sec:classical}
        To understand the quantum regime of cavity optomechanics in a better way, let us first explore a classical model of optomechanical systems.
        Such a description can be obtained from the QLEs in Equations \eqref{eqn:oms_qle_eom} by analyzing the complex field amplitudes $\alpha = \langle a \rangle$ and $\beta = \langle b \rangle$.
        Utilizing the time-average properties of the noises, we obtain the coupled set of equations
        \begin{subequations}
            \label{eqn:classical_dynamics}
            \begin{eqnarray}
                \dot{\alpha} & = & - \left( \frac{\kappa}{2} - i \Delta \right) \alpha + A_{l}, \\
                \dot{\beta} & = & - \left( \frac{\gamma}{2} + i \omega_{m} \right) \beta + i g_{0} \alpha^{*} \alpha,
            \end{eqnarray}
        \end{subequations}
        where $\Delta = \Delta_{0} + g_{0} ( \beta^{*} + \beta )$ is the \textit{effective detuning}.
        
        The classical equations of motion described in Equations \eqref{eqn:classical_dynamics} lead to some interesting observations.
        To analyze these, let us first assume that the optical and mechanical modes only deviate slightly from their steady-state values.
        This lets us linearize the dynamics around the steady state and study the effect of the classical deviations on the characteristics of the optical and mechanical modes.

        \subsection{Steady State and Bistability}
            \label{sec:classical_bistability}
            In the long-time limit, the modes can be thought of as entering a steady-state such that their derivatives $d \langle \mathcal{O} \rangle / d t$ are approximately equal to zero. 
            We then obtain the steady-state mode amplitudes as
            \begin{subequations}
                \label{eqn:classical_bistability_steady}
                \begin{eqnarray}
                    \alpha_{s} & = & \frac{A_{l}}{\frac{\kappa}{2} - i \Delta}, \\
                    \beta_{s} & = & \frac{i g_{0} \left| \alpha_{s} \right|^{2}}{\frac{\gamma}{2} + i \omega_{m}}.
                \end{eqnarray}
            \end{subequations}

\begin{figure}[!ht]
                \centering
                \includegraphics[width=0.4\textwidth]{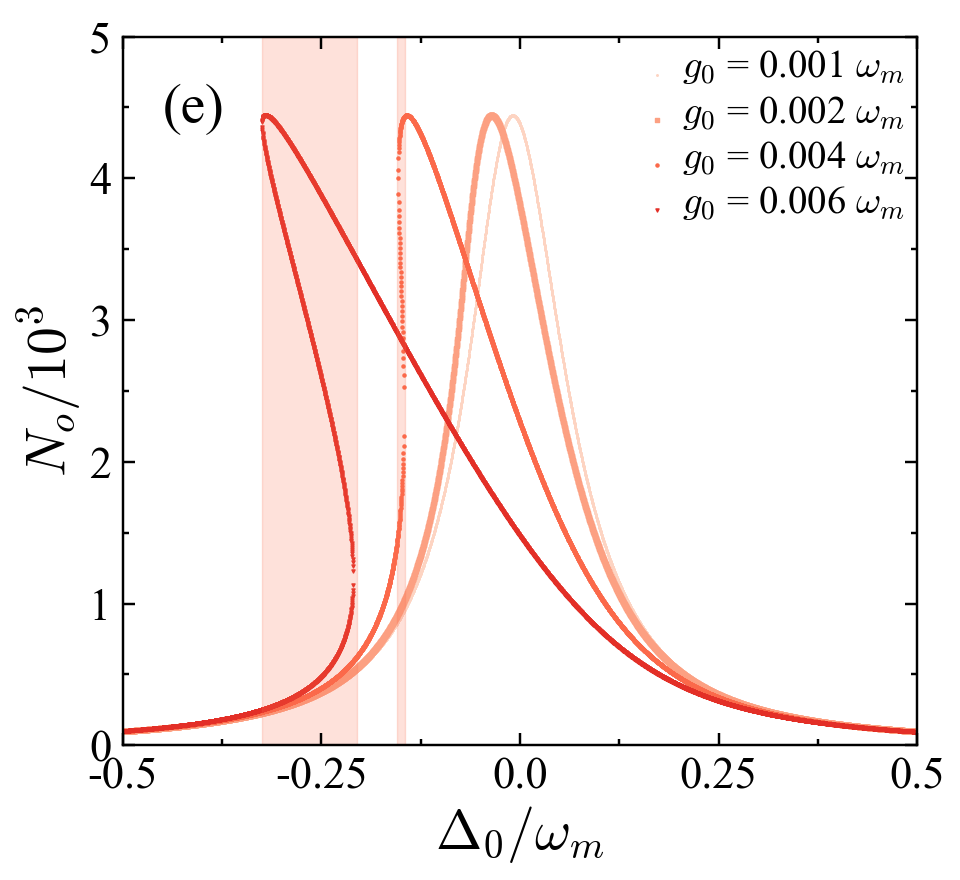}
                \includegraphics[width=0.4\textwidth]{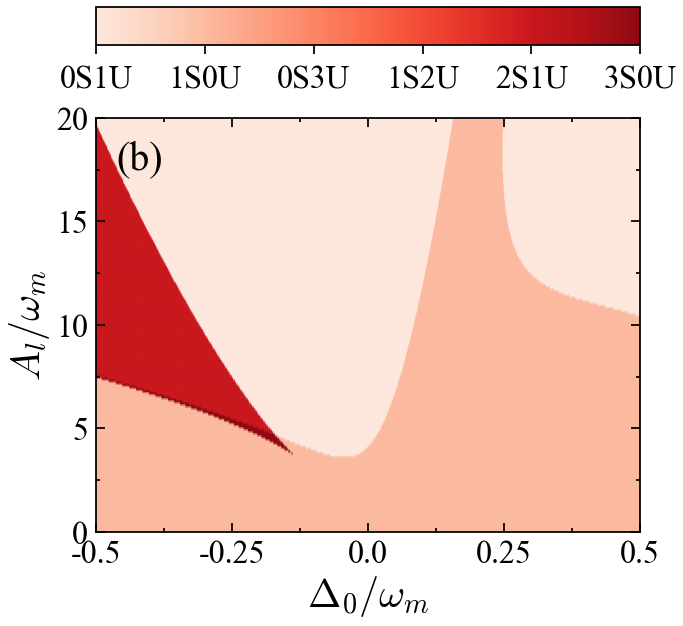}
                \caption{
                    (a) Number of intracavity photons ($N_{o}$) with variation in the laser detuning ($\Delta_{0}$) for different values of optomechanical coupling ($g_{0}$).
                    The shaded areas denote the regions of bistability.
                    (b) Stability of the system for different values of the laser detuning ($\Delta_{0}$) and laser amplitude ($A_{l}$) calculated using the Routh-Hurwitz criteria.
                    Here `S' denotes a stable branch whereas `U' denotes an unstable branch.
                    The parameters used for the plots (in units of $\omega_{m}$) are $A_{l} = 5.0$, $\gamma = 0.005$ and $\kappa = 0.15$.
                }
                \label{fig:classical_bistability}
            \end{figure}

            Equations \eqref{eqn:classical_bistability_steady} give rise to a cubic equation in the \textit{mean optical occupancy} $N_{o} = | \alpha_{s} |^{2}$, marking two categories of steady-state solutions.
            One solution has one real root while the other one has three real roots.  
            The latter case is the multistable steady-state regime (shaded regions in Figure \ref{fig:classical_bistability} (a)).
            However, the middle branch of solutions is not traversed while varying the detuning and the system displays discontinuous jumps between the upper and lower states displaying a hysteretic behaviour \cite{PhysRevLett.51.1550}. 
            As such, in the second regime, the system can be treated as \textit{effectively bistable} \cite{TaylorFrancis.QuantumOptomechanics.Bowen}.
            \\

            \textbf{Exercise 4:} Derive the cubic equation in $N_{o}$ and verify the bistability curve in Figure \ref{fig:classical_bistability} (a).
            \\

            One can also analyze the stability of the obtained steady-state values of $\alpha_{s}$ from the cubic in $N_{o}$ by utilizing the Routh-Hurwitz criteria for nonlinear ordinary differential equations \cite{PhysRevA.35.5288}.
            Such an examination is usually performed to locate the monostable branches of the system and drive the system away from instability \cite{NewJPhys.22.013049}.
            A typical plot for stable and unstable branches is shown in Figure \ref{fig:classical_bistability} (b).

        \subsection{Response to an External Force}
            \label{sec:classical_response}
            In order to study the linear response of the system to an external force $F_{ext}$ acting on the mirror, we write the classical deviations in the optical mode amplitude and the mechanical displacement as $\delta \alpha = \alpha - \alpha_{s}$ and $\delta q = q - q_{s}$ ($q = \langle x \rangle$) respectively.
            In the frequency domain, the corresponding coupled equations can be obtained as
            \begin{subequations}
                \label{eqn:classical_response_frequency}
                \begin{eqnarray}
                    \left( - i \omega + \frac{\kappa}{2} - i \Delta \right) \delta \alpha ( \omega ) = i G \alpha_{s} q (\omega), \\
                    \left( - m \omega^{2} + m \omega_{m}^{2} - i m \gamma \omega \right) q (\omega) \nonumber \\
                    = \hbar G \left( \alpha_{s} \delta \alpha^{*} (- \omega) + \alpha_{s}^{*} \delta \alpha ( \omega ) \right) + F_{ext} (\omega),
                \end{eqnarray}
            \end{subequations}
            where we have used the effective detuning $\Delta = \Delta_{0} + G q_{s}$ ($q_{s}$ is the steady-state displacement) and the fact that $\mathcal{F} [ \delta \alpha^{*} (t) ] = \delta \alpha^{*} (- \omega)$.
            \\

            \textbf{Exercise 5:} Derive Equations \eqref{eqn:classical_response_frequency}.
            \textit{Hint:} Use the relations $x = x_{ZP} ( b^{\dagger} + b )$ and $p = i p_{ZP} ( b^{\dagger} - b )$.
            \\

            Now, the first equation can be written as $\delta \alpha (\omega) = i G \alpha_{s} \chi_{o} (\omega) q$, where 
            \begin{equation}
                \label{eqn:classical_response_chi_o}
                \chi_{o} (\omega) = \frac{1}{\frac{\kappa}{2} - i \left( \Delta + \omega \right)}
            \end{equation}    
            is the \textit{optical susceptibility}, which relates the variation of the optical amplitude with variation in the mechanical motion.
            A similar expression for mechanical susceptibility can be obtained in the absence of optomechanical coupling as
            \begin{equation}
                \label{eqn:classical_response_chi_m}
                \chi_{m} (\omega) = \frac{1}{m \left( \omega_{m}^{2} - \omega^{2} \right) - i m \gamma \omega},
            \end{equation}
            such that $q (\omega) = \chi_{m} (\omega) F_{ext} (\omega)$ relates the external force with the variation in mechanical displacement.

            In the presence of optomechanical coupling, the mechanical susceptibility gets modified to
            \begin{equation}
                \label{eqn:classical_response_chi}
                \chi (\omega) = \frac{1}{\frac{1}{\chi_{m} (\omega)} - \Sigma (\omega)},
            \end{equation}
            where the \textit{optomechanical self energy} is defined as
            \begin{equation}
                \label{eqn:classical_response_sigma}
                \Sigma (\omega) = 2 i m \omega_{m} g_{s}^{2} \left( \chi_{o} (\omega) - \chi_{o}^{*} (- \omega) \right),
            \end{equation}
            with $g_{s} = G x_{ZP} | \alpha_{s} | = g_{0} | \alpha_{s} |$.
            The introduction of $\Sigma (\omega)$ in the mechanical response leads to two characteristic phenomena in a cavity optomechanical system as discussed next.

        \subsection{Optomechanical Damping}
            \label{sec:classical_damping}
            From the second equation of Equation \eqref{eqn:classical_response_frequency}, it is evident that the imaginary part of $\Sigma (\omega \approx \omega_{m}) / ( m \omega_{m} )$ gives the \textit{optomechanical damping rate},
            \begin{equation}
                \label{eqn:classical_damping}
                \gamma_{om} = g_{s}^{2} \kappa \left\{ \frac{1}{\frac{\kappa^{2}}{4} + \left( \omega_m + \Delta \right)^{2}} - \frac{1}{\frac{\kappa^{2}}{4} + \left( \omega_{m} - \Delta \right)^{2}} \right\}.
            \end{equation}

            The effective mechanical damping rate can then be expressed as a sum of the original mechanical damping rate $\gamma$ and this new rate as discussed in Section \ref{sec:basic_dynamics}.
            Thus, depending on the value of $\gamma_{om}$, the mechanical motion can either encounter cooling or heating.
            \begin{figure}[h!]
                \centering
                \includegraphics[width=0.4\textwidth]{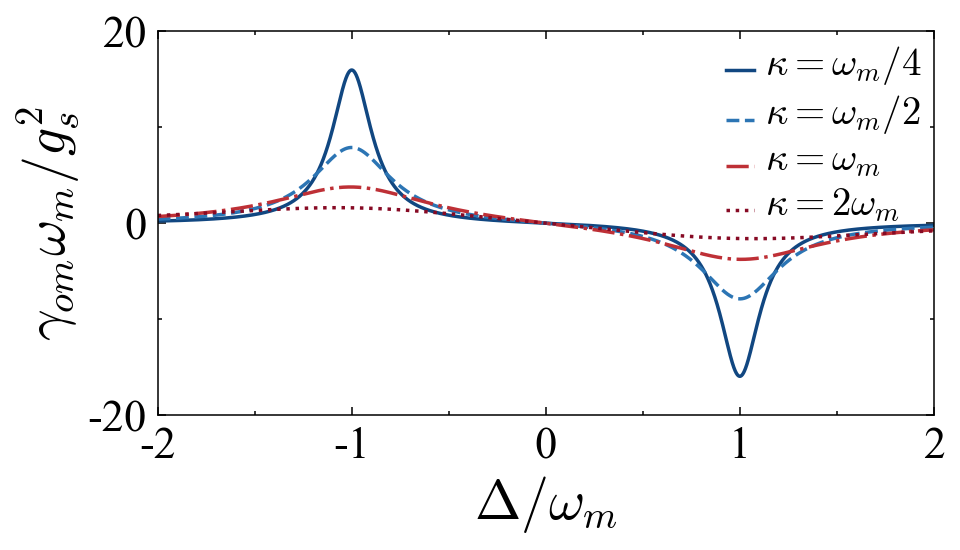}
                \caption{Optomechanical damping ($\gamma_{om}$) as a function of the effective detuning ($\Delta$).}
                \label{fig:classical_damping}
            \end{figure}

            In Figure \ref{fig:classical_damping}, we plot $\gamma_{om}$ as a function of the effective detuning $\Delta$.
            It can be seen that when the cavity decay rate is much smaller than the mechanical frequency, i.e., $\kappa \ll \omega_{m}$, the total damping increases around $\Delta = - \omega_{m}$.
            This cools down the mechanical oscillator.
            Whereas, around $\Delta = \omega_{m}$, the damping decreases and the mechanical oscillator heats up.
            The first situation also corresponds to the resonant absorption of a phonon to accommodate a blue-shifted photon inside the cavity, thereby cooling the mechanical oscillator.
            On the other hand, the second scenario may lead to the excitation of a phonon with a red-shifted photon inside the cavity.
            The regime where $\kappa \ll \omega_{m}$ is known as the \textit{resolved sideband regime}.
            Here, the linewidth of the cavity mode is small enough to resolve the blue-shifted or red-shifted photons exiting the cavity.
            \\

            \textbf{Exercise 6:} What happens when $\gamma_{om} < - \gamma$?
            \\

        \subsection{Optical Spring Effect}
            \label{sec:classical_spring}
            It can also be seen that the real part of $\Sigma (\omega \approx \omega_{m}) / ( 2 m \omega_{m} )$ is related to the \textit{frequency shift} in the mechanical mode induced by the optical field,
            \begin{equation}
                \label{eqn:classical_spring}
                \delta \omega_{m} = g_{s}^{2} \left\{ \frac{\omega_{m} + \Delta}{\frac{\kappa^{2}}{4} + \left( \omega_{m} + \Delta \right)^{2}} - \frac{\omega_{m} - \Delta}{\frac{\kappa^{2}}{4} + \left( \omega_{m} - \Delta \right)^{2}} \right\}.
            \end{equation}

            It can be seen that the mechanical oscillator's response is either enhanced or slowed down by the optical field.
            This phenomena where the mechanical spring constant is altered by the optical field is known as the \textit{optical spring effect}.
            \begin{figure}[h!]
                \centering
                \includegraphics[width=0.4\textwidth]{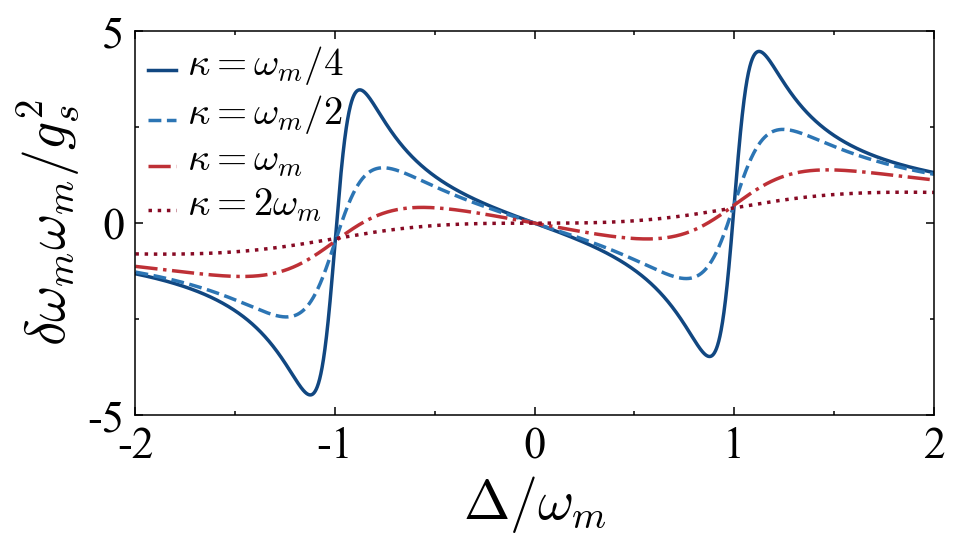}
                \caption{Mechanical frequency shift ($\delta \omega_{m}$) as a function of the effective detuning ($\Delta$).}
                \label{fig:classical_spring}
            \end{figure}

            Figure \ref{fig:classical_spring} shows a typical plot of the frequency shift as a function of the effective detuning.
            When $\kappa \gg \omega_{m}$, the spring gets stiffer when $\Delta > 0$, i.e., when the laser is blue-detuned.
            On the other hand, when $\Delta < 0$, i.e., when the laser is red-detuned, the spring gets softer.
            However, if we work in the resolved sideband regime $\kappa \ll \omega_{m}$, the stiffening or softening of the spring can be appropriately adjusted by tuning the laser near $| \Delta | = \omega_{m}$.
            \\
            
            \textbf{Exercise 7:} What happens when $\delta \omega_{m} < - \omega_{m}$ in the red-detuned regime?
            \\


    \section{Quantum Optomechanics}
        \label{sec:quant}
        
        In Section \ref{sec:oms_qle}, we mentioned that optomechanical cavities are typical open quantum systems.
        A cavity optomechanical system is easy to deal quantum mechanically as the Hamiltonian describing such systems are akin to the so-called Jaynes Cummings model \cite{RevModPhys.86.1391}.
        Below, we give a brief description of the linearized cavity quantum optomechanics. 
        
        \subsection{Equations of Fluctuations}
            \label{sec:quantum_equations}
            When the cavity is strongly driven, the radiation pressure force enhances considerably due to the large number of intracavity photons.
            In such a scenario, even for small values of optomechanical coupling, the system can be approximated by a linearized description.
            This means that the mode operators can be expressed as a sum of their classical mean amplitudes and the quantum fluctuations around these classical values, that is, $a = \alpha + \delta a$ and $b = \beta + \delta b$.
            The coupled equations for the quantum fluctuations can therefore be obtained from Equations \eqref{eqn:oms_qle_eom} as,
            \begin{subequations}
                \label{eqn:quantum_equations}
                \begin{eqnarray}
                    \delta \dot{a} & = & - \left( \frac{\kappa}{2} - i \Delta \right) \delta a + i g \left( \delta b^{\dagger} + \delta b \right) \nonumber \\
                    && + \sqrt{\kappa} a_{in}, \\
                    \delta \dot{b} & = & - \left( \frac{\gamma}{2} + i \omega_{m} \right) \delta b + i \left( g \delta a^{\dagger} + g^{*} \delta a \right) \nonumber \\
                    && + \sqrt{\gamma} b_{in}.
                \end{eqnarray}
            \end{subequations}

            Here, $\Delta = \Delta_{0} + g_{0} ( \beta^{*} + \beta )$ is the effective detuning defined earlier, and $g = g_{0} \alpha$ is known the \textit{effective optomechanical coupling constant} \cite{TaylorFrancis.QuantumOptomechanics.Bowen}.
            Thus, we see that the optomechanical coupling strength is enhanced by the optical mode amplitude.
            It may also be noted that since the amplitude of fluctuations is much smaller than their classical values, we have ignored the second-order terms in $\delta a$ and $\delta b$ under the linearized approximation.\\

            \textbf{Exercise 8:} Derive Equations \eqref{eqn:quantum_equations}.
            \\

            One can also define the fluctuation quadratures for the optical mode ($\delta X = (\delta a^{\dagger} + \delta a) / \sqrt{2}$, $\delta Y = i (\delta a^{\dagger} - \delta a) / \sqrt{2}$) and the mechanical mode, ($\delta Q = (\delta b^{\dagger} + \delta b) / \sqrt{2}$, $\delta P = i (\delta b^{\dagger} - \delta b) / \sqrt{2}$), and write a compact form of the dynamics as,
            \begin{equation}
                \label{eqn:quantum_linear_compact}
                \dot{{\mathbf{u}}} (t) = \mathbf{A} (t) \mathbf{u} (t) + \mathbf{n} (t),
            \end{equation}
            where $\mathbf{u} = {(\delta X, \delta Y, \delta Q, \delta P)}^{T}$, $\mathbf{A}$ is known as the \textit{drift} matrix, and $\mathbf{n}$ is the vector containing the noises induced in the quadratures.
            \\

            \textbf{Exercise 9:} Obtain the expressions for the drift matrix $\mathbf{A}$ and the noise vector $\mathbf{n}$.
            \\

        \subsection{Linearized Hamiltonian}
            \label{sec:quantum_linear}
            The Hamiltonian describing the dynamics of the fluctuations in Equations \eqref{eqn:quantum_equations} can be written as \cite{TaylorFrancis.QuantumOptomechanics.Bowen}
            \begin{eqnarray}
                \label{eqn:quantum_linear_h}
                H_{lin} & = & - \hbar \Delta \delta a^{\dagger} \delta a + \hbar \omega_{m} \delta b^{\dagger} \delta b \nonumber \\
                && - \hbar \left( g \delta a^{\dagger} + g^{*} \delta a \right) \left( \delta b^{\dagger} + \delta b \right).
            \end{eqnarray}
            \begin{figure}[h!]
                \centering
                \includegraphics[width=0.48\textwidth]{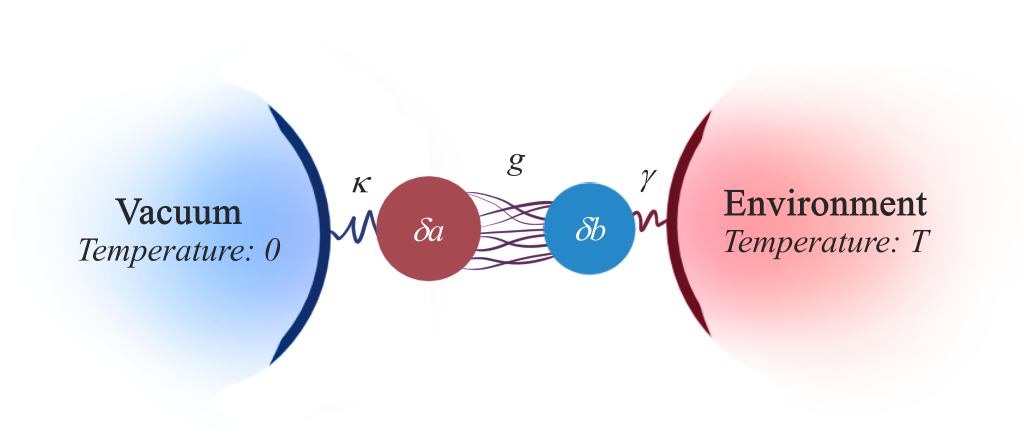}
                \caption{
                    An illustration of the linearized optomechanical interaction.
                    The cavity mode is assumed to be connected to a vacuum bath whereas the mechanical mode is in contact with a thermal environment.
                    Source: \cite{AVSQuantumSci.3.015901}.
                }
                \label{fig:quantum_linear}
            \end{figure}

            A pictorial depiction of this linearized interaction is shown in Figure \ref{fig:quantum_linear}.
            The \textit{linearized} Hamiltonian also explains a variety of interesting physical phenomena reported in optomechanical systems.
            Let us briefly analyze these.
            Assuming a simple scenario where the classical values settle down to a steady state and the phase of the laser light is adjusted in such a way that $\alpha$ is real, i.e., $g = g_{0} \alpha = g_{0} | \alpha_{s} | = g_{s}$, we obtain a simplified Hamiltonian
            \begin{eqnarray}
                \label{eqn:quantum_linear_simple}
                H_{lin} & = & - \hbar \Delta \delta a^{\dagger} \delta a + \hbar \omega_{m} \delta b^{\dagger} \delta b \nonumber \\
                && - \hbar g_{s} \left( \delta a^{\dagger} + \delta a \right) \left( \delta b^{\dagger} + \delta b \right).
            \end{eqnarray}
            
            It can be clearly seen that this Hamiltonian features a position-position optomechanical interaction which is enhanced by the effective coupling constant.
            It is also evident from Equations \eqref{eqn:quantum_equations} that the displacement of the mechanical oscillator is proportional to the phase shift of the cavity field. 
            Therefore, measurement of the phase of the output field of the cavity provides information on the position of the moveable mirror and this facilitates the \textit{readout} of the mechanical position.
            For details on readout and measurement, the readers may refer to \cite{RevModPhys.86.1391, TaylorFrancis.QuantumOptomechanics.Bowen, SBH.CavityOptomechanics.Aspelmeyer, PhysRevLett.70.548}.

            When the cavity is driven, the light scattered by the mechanical mirror gives rise to Stokes and anti-Stokes sidebands where a quanta of mechanical energy is either absorbed or excited by the light field respectively.
            Depending on the detuning of the laser, this leads to either cooling or heating as discussed in the previous sections.
            These processes can be mathematically visualized by writing the Hamiltonian of Equation \eqref{eqn:quantum_linear_simple} in the interaction picture as
            \begin{eqnarray}
                \label{eqn:quantum_linear_int}
                H_{int} & = & - \hbar g_{s} \Bigg[ \left\{ \delta a^{\dagger} \delta b e^{- i \left( \Delta + \omega_{m} \right)} + \delta a \delta b^{\dagger} e^{i \left( \Delta + \omega_{m} \right)} \right\} \nonumber \\
                && + \left\{ \delta a^{\dagger} \delta b^{\dagger} e^{- i \left( \Delta - \omega_{m} \right)} + \delta a \delta b e^{i \left( \Delta - \omega_{m} \right)} \right\} \Bigg].
            \end{eqnarray}
            \\

            \textbf{Exercise 10:} Derive Equation \eqref{eqn:quantum_linear_int}.
            \\

            Under the rotating wave approximation (RWA) \cite{TaylorFrancis.QuantumOptomechanics.Bowen}, we obtain the following forms of interactions that lead to the two distinct phenomena of cooling and heating.

            \paragraph{Beam-splitter Interaction and Cooling:}
                In the red-detuned regime, when $\Delta = - \omega_m$, the interaction Hamiltonian under RWA takes the form of a \textit{beam-splitter interaction}
                \begin{equation}
                    \label{eqn:quantum_linear_cooling}
                    H_{int} = - \hbar g_{s} \left( \delta a^{\dagger} \delta b + \delta a \delta b^{\dagger} \right).
                \end{equation}
                Since the mechanical mode is at a higher temperature than the optical mode, such a Hamiltonian preserves the phonon de-excitation process and this leads to a \textit{cooling} effect in the mechanical motion.
                Cooling of the mechanical mode is a requirement for most quantum optomechanical applications and various schemes have been proposed to cool the mechanical oscillator to its quantum ground state \cite{PhysRevLett.99.093901, NatPhys.4.415}. 

            \paragraph{Squeezed Interaction and Heating:}
                Similarly, when $\Delta = \omega_{m}$ in the blue-detuned regime, the Hamiltonian takes a \textit{squeezing interaction} form
                \begin{equation}
                    \label{eqn:quantum_linear_heat}
                    H_{int} = - \hbar g_{s} \left( \delta a^{\dagger} \delta b + \delta a \delta b^{\dagger} \right).
                \end{equation}
                However, the resonant interaction of the modes supresses the photon de-excitation process and this leads to \textit{heating} or \textit{amplification} of the mechanical mode through simultaneous excitation of an optical and mechanical quanta.
                Additionally, this form can also be used to squeeze the variance of the position fluctuations ($\langle Q^{2} \rangle$) of the mechanical mode at the expense of increased variance in the momentum fluctuations ($\langle P^{2} \rangle$) \cite{PhysRevLett.103.213603, PhysRevA.101.053836}.
                This is routinely used in interferometric setups to enhance the sensitivity of the mechanical motion \cite{RepProgPhys.72.076901, PhysRevLett.123.231108}.
                
        \subsection[Macroscopic Quantum Phenomena]{Macroscopic Quantum \\Phenomena}
            \label{sec:quantum_phenomena}
            Apart from phenomena like ground state cooling and mechanical squeezing which are used to prepare highly-sensitive mechanical oscillators, optomechanical systems also provide a suitable testbed for the study of quantum phenomena in the macroscopic domain.
            An important ingredient in the calculation of most of the figure of merits for these phenomena is the \textit{correlation matrix} \cite{PhysRevLett.98.030405}
            \begin{equation}
                \label{eqn:quantum_phenomena_v}
                \mathbf{V}_{ij} = \frac{1}{2} \langle \mathbf{u}_{i} \mathbf{u}_{j} + \mathbf{u}_{j} \mathbf{u}_{i} \rangle.
            \end{equation}

            Using the formal solution for Equation \eqref{eqn:quantum_linear_compact} as $\mathbf{u}_{i} (t) = \sum_{k} \{ \mathbf{M}_{ik} (t, t_{0}) \mathbf{u}_{k} (t_{0}) + \int_{t_{0}}^{t} ds \mathbf{M}_{ik} (t, s) \mathbf{n}_{k} (s) \}$, such that $\dot{\mathbf{M}} (t, t_{0}) = \mathbf{A} (t) \mathbf{M} (t, t_{0})$ with $\mathbf{M} (t, t) = \mathbb{1}$, it can be shown that the correlation matrix follows the dynamical equation of motion
            \begin{equation}
                \label{eqn:quantum_phenomena_dynamics}
                \dot{\mathbf{V}} (t) = \mathbf{A} (t) \mathbf{V} (t) + \mathbf{V} (t) \mathbf{A}^{T} (t) + \mathbf{D},
            \end{equation}
            where the noise correlations are contained in $\mathbf{D} = \mathrm{diag} [ \kappa / 2, \kappa / 2, \gamma ( n_{th} + 1 /2 ), \gamma ( n_{th} + 1 /2 ) ]$.
            \\
            
            \textbf{Exercise 11:} Derive Equation \eqref{eqn:quantum_phenomena_dynamics}.
            \\

            It is important to note here that the stability of the formal solution is necessary to obtain the dynamical form of the correlation matrix.
            This can be done by analyzing the eigenvalues of $A$ through the Routh-Hurwitz criteria introduced in Section \ref{sec:classical_bistability}.
            With this, let us briefly introduce three quantum phenomena that are being extensively studied in recent years.
            The inquisitive reader may refer to the cited articles for further insights on the mathematical formalism of each of these phenomena.

            \paragraph{State-transfer:}
                Since mechanical decays are much smaller than that of their optical counterparts, encoding the optical state in the mechanical mode can be done to store information for longer durations.
                Also, a common mechanical mode can be used to transduce information between optical and microwave signals \cite{NewJPhys.14.105010}.
                Schemes to transfer the optical state from one node of a many-body optomechanical system to another has applications in the design of quantum networks \cite{PhysRevA.93.062339}.
                Transduction between optical, mechanical and microwave domains is currently an active field of research \cite{Nature.588.599, PhysRevA.92.043845, NatPhys.10.321}.

            \paragraph{Entanglement:}
                Entanglement is a extremely important resource for quantum information processing and quantum communication \cite{CUP.QuantumComputationQuantumInformation.Nielsen, CUP.QuantumInformationTheory.Wilde}.
                The quantification of entanglement between the optical and mechanical elements of an optomechanical system was pioneered by Vitali \textit{et. al.} in 2008 \cite{PhysRevLett.98.030405}.
                Since then, numerous schemes have been proposed to strongly entangle optical and mechanical components of single as well as many-body optomechanical systems \cite{AVSQuantumSci.3.015901, PhysRevA.86.042306, PhysRevA.94.053807}.
                Entangling microwave and optical signals to create hybrid electro-optomechanical converters can find applications in the development of quantum sensors and radars in the radio-frequency domain \cite{NatNanotechnol.13.11, PhysRevA.84.042342, PhysRevLett.114.080503, NewJPhys.22.063041, PhysRevA.106.043501}.

            \paragraph{Synchronization:}
                In the classical world, synchronization \textemdash{} a tendency of oscillators to sympathetically adjust their rhythms \textemdash{} is a ubiquitous phenomena observed across different domains \cite{PBL.Sync.Strogatz}.
                Several measures of quantum synchronization have also been proposed in recent years as this phenomena is gathering a good amount of interest in the deep quantum regime \cite{SIP.QuantumCorrSyncMeasures.Glave, PhysRevA.91.012301, PhysRevA.99.043804, PhysRevLett.111.103605, PhysRevResearch.2.043287}.
                Since an optomechanical system can undergo self-oscillations, coupled optomechanical systems can also display synchronization \cite{PhysRevLett.109.233906, PhysRevLett.111.103605, PhysRevA.99.033818}.
                Long-distance quantum synchronization can find applications in quantum communication and quantum networks \cite{SciRep.3.01439, OptExpress.24.012336, CommunNonlinearSciNumerSimulat.42.121, PhysRevE.95.022204}.
                
    \section{Conclusion and Future Prospects}
        \label{sec:conclusion}
        In this tutorial, we introduced an optomechanical system with an optomechanical coupling of the first order.
        We then systematically derived the classical and quantum dynamics for such a system, detailing on some characteristic properties and phenomena.
        We believe that the theories discussed will help the reader to get started in the field of optomechanics and analyze a variety of phenomena for linearizable quantum optomechanical systems.

        In the last two decades, studies have also explored the effect of higher-order coupling terms and noises \cite{Photonics.4.48, SciRep.8.16676, PhysRevA.82.021806, NewJPhys.18.013043, PhysRevA.96.043832, PhysRevE.83.056202}, photon blockade \cite{PhysRevA.88.023853, PhysRevA.98.013826}, induced transparency \cite{PhysRevA.81.041803, PhysRevA.88.013804, PhysRevA.101.043820} and collective nonlinear dynamics \cite{PhysRevLett.107.043603, Nature.600.75}.
        It is also inspiring to see emerging interdisciplinary studies that utilize the formalism of optomechanics \cite{NatPhys.18.15, RevModPhys.93.025005, PhysRevLett.127.113601, FrontPhys.17.42201}.
        We believe that the field of optomechanics shall serve as a versatile platform to propose, interpret or predict quantum phenomena as well as advance quantum technologies in the years to come.

    \section*{Acknowledgement}
        S. K. would like to acknowledge MHRD, Government of India for providing financial support for his research through the PMRF scheme.

    \appendix

    \section*{Solutions to Exercises}
    
        \textbf{Exercise 1:} Obtain the radiation pressure force and its relationship with the frequency pull parameter using the Hamiltonian in Equation \eqref{eqn:oms_h_om}.

        \textbf{Solution:} It can be easily derived from Equation \eqref{eqn:oms_h_om} that the frequency pull parameter is also related to the radiation pressure force $\hat{F}_{RP}$ as
        \begin{equation}
            \label{eqn:oms_h_rpf}
            \hat{F}_{RP} = - \frac{\delta \hat{H}_{om}}{\delta \hat{q}} = \frac{\hbar \omega_{o} \hat{n}_{a}}{L} = \hbar G \hat{a}^{\dagger} \hat{a}. \nonumber
        \end{equation}
        
        \textbf{Exercise 2:} Choose an appropriate unitary transformation and derive Equation \eqref{eqn:oms_h_l}.

        \textbf{Solution:} We start with the Hamiltonian defined in Equation \eqref{eqn:oms_h}, and the Schr\"{o}dinger equation $i \hbar d \ket{\Psi_{S}} / dt = \hat{H} \ket{\Psi_{S}}$, and define a unitary operator $U$ such that $\ket{\Psi_{I}} = \hat{U} \ket{\Psi_{S}}$ is the new wavefunction in the interaction picture, $\ket{\Psi_{S}}$ being the one in the Schr\"{o}dinger picture.
        Substituting the new wavefunction with $\hat{U} = \mathrm{exp} [ i \omega_{l} \hat{a}^{\dagger} \hat{a} t ]$ and $\hat{U} \hat{U}^{\dagger} = \mathbb{1}$, we get,
        \begin{equation}
            i \hbar \frac{d \ket{\Psi_{I}}}{dt} = \left( \hat{U} \hat{H} \hat{U}^{\dagger} - \hbar \omega_{l} \hat{a}^{\dagger} \hat{a} \right) \ket{\Psi_{I}} = \hat{H}_{I} \ket{\Psi_{I}}. \nonumber
        \end{equation}

        Using the Baker-Campbell-Hausdroff formula, 
        \begin{eqnarray}
            e^{\hat{A}} \hat{B} e^{\hat{A}} & = & \hat{B} + \left[ \hat{A}, \hat{B} \right] + \frac{1}{2!} \left[ \hat{A}, \left[ \hat{A}, \hat{B} \right] \right] + \dots \nonumber \\
            && + \frac{1}{n!} \left[ \hat{A}, \left[ \hat{A}, \dots \left[ \hat{A}, \hat{B} \right] \dots \right] \right] + \dots, \nonumber
        \end{eqnarray}
        along with the Bosonic commutation relations for $\hat{a}$ and $\hat{b}$, we obtain the relations
        \begin{subequations}
            \begin{eqnarray}
                \hat{U} \hat{a}^{\dagger} \hat{U}^{\dagger} = \hat{a}^{\dagger} e^{i \omega_{l} t}, \nonumber \\
                \hat{U} \hat{a} \hat{U}^{\dagger} = \hat{a} e^{- i \omega_{l} t}, \nonumber
            \end{eqnarray}
        \end{subequations}
        which gives us in the Hamiltonian of Equation \eqref{eqn:oms_h_l}.
        \\

        \textbf{Exercise 3:} Using the Bosonic commutation relations, derive Equations \eqref{eqn:oms_qle_eom}.

        \textbf{Solution:} Commutation of the Hamiltonian $H_{sys}$ in Equation \eqref{eqn:oms_h_l} with $a$ gives us
        \begin{equation}
            \frac{1}{i \hbar} \left[ a, H_{sys} \right] = i \Delta_{0} a + i g_{0} \left( b^{\dagger} + b \right) a + A_{l}. \nonumber
        \end{equation}

        Similarly, commutation with $b$ gives us
        \begin{equation}
            \frac{1}{i \hbar} \left[ b, H_{sys} \right] = - i \omega_{m} a + i g_{0} a^{\dagger} a. \nonumber
        \end{equation}

        Together with the input noises and the decay terms, these equations results in Equation \eqref{eqn:oms_qle_eom}.
        \\

        \textbf{Exercise 4:} Derive the cubic equation in $N_{o}$ and verify the bistability curve in Figure \ref{fig:classical_bistability} (a).

        \textbf{Solution:} Eliminating the mechanical mode from the steady-state amplitudes derived from Equations \eqref{eqn:classical_dynamics} in the adiabatic limit, we obtain a cubic equation for the optical field occupancy number $N_{o} = | \alpha_{s} |^{2}$, which reads as 
        \begin{equation}
            N_{o} \left\{ \frac{\kappa^{2}}{4} + \left( \Delta_{0} + C N_{o} \right)^{2} \right\} = \left| A_{l} \right|^{2}, \nonumber
        \end{equation}
        where $C = 2 g_{0}^{2} \omega_{m} / ( \gamma^{2} / 4 + \omega_{m}^{2} )$.
        It can further be simplified to
        \begin{eqnarray}
            4 C^{2} N_{o}^{3} + 8 C \Delta_{0} N_{o}^{2} + \left( 4 \Delta_{0}^{2} + \kappa^{2} \right) N_{o} \nonumber \\
            - 4 \left| A_{l} \right|^{2} = 0. \nonumber
        \end{eqnarray}

        \textbf{Exercise 5:} Derive Equations \eqref{eqn:classical_response_frequency}.
        \textit{Hint:} Use the relations $x = x_{ZP} ( b^{\dagger} + b )$ and $p = i p_{ZP} ( b^{\dagger} - b )$.
        
        \textbf{Solution:} We write the Hamiltonian in Equation \eqref{eqn:oms_h_l} in terms of the operators for position $x = \sqrt{\hbar / ( 2 m \omega_{m} )} ( b^{\dagger} + b )$ and momentum $p = i \sqrt{\hbar m \omega_{m} / 2} ( b^{\dagger} - b )$ as
        \begin{eqnarray}
            H_{sys} & = & - \hbar \Delta_{0} a^{\dagger} a + \frac{p^{2}}{2 m} + \frac{m \omega_{m}^{2} x^{2}}{2} - \hbar G a^{\dagger} a x \nonumber \\
            && + i \hbar A_{l} (a^{\dagger} - a). \nonumber
        \end{eqnarray}

        The corresponding QLEs are obtained as
        \begin{subequations}
            \begin{eqnarray}
                \dot{a} & = & - \left\{ \frac{\kappa}{2} - i \left( \Delta_{0} + G x \right) \right\} a \nonumber \\
                && + A_{l} + \sqrt{\kappa} a_{in}, \nonumber \\
                \dot{x} & = & \frac{p}{m}, \nonumber \\
                \dot{p} & = & - \gamma p - m \omega_{m}^{2} x + \hbar G a^{\dagger} a + \zeta_{in}, \nonumber
            \end{eqnarray}
        \end{subequations}
        where $\zeta_{in}$ contains the fluctuations entering into the mechanical momentum.
        The classical mean amplitudes $\alpha = \langle a \rangle$ and $q = \langle x \rangle$ can then be obtained as
        \begin{subequations}
            \begin{eqnarray}
                \dot{\alpha} = - \left\{ \frac{\kappa}{2} - i \left( \Delta_{0} + G q \right) \right\} \alpha + A_{l}, \nonumber \\
                m \ddot{q} + m \omega_{m}^{2} q + m \gamma \dot{q} = \hbar G | \alpha |^{2}, \nonumber
            \end{eqnarray}
        \end{subequations}
        In the long-time limit, the steady-state value for the optical amplitude can be obtained as 
        \begin{subequations}
            \begin{eqnarray}
                \alpha_{s} & = & \frac{A_{l}}{\frac{\kappa}{2} - i \Delta}, \nonumber \\
                q_{s} & = & \frac{\hbar G | \alpha |^{2}}{m \omega_{m}^{2}}, \nonumber
            \end{eqnarray}
        \end{subequations}
        where $\Delta = \Delta_{0} + G q_{s}$  is the steady-state value of the mechanical displacement.
        Now, we introduce the classical deviations around the mean steady-state amplitudes as $\delta \alpha = \alpha - \alpha_{s}$ and $\delta q = q - q_{s}$.
        Ignoring the second order terms of the deviations, we get
        \begin{subequations}
            \begin{eqnarray}
                \delta \dot{\alpha} = - \left( \frac{\kappa}{2} - i \Delta \right) \delta \alpha + i G \alpha_{s} \delta q, \nonumber \\
                m \delta \ddot{q} + m \omega_{m}^{2} \delta q + m \gamma \delta \dot{q} \nonumber \\
                = \hbar G \left( \alpha_{s} \delta \alpha^{*} + \alpha_{s}^{*} \delta \alpha \right). \nonumber
            \end{eqnarray}
        \end{subequations}
        Upon Fourier transform, these equations give Equation \eqref{eqn:classical_response_frequency}.
        \\

        \textbf{Exercise 6:} What happens when $\gamma_{om} < - \gamma$?
        
        \textbf{Answer:} An interesting phenomena is observed when $\gamma_{om} < - \gamma$.
        As the total mechanical damping becomes negative, the system experiences gain and the nonlinear effects lead to self-induced oscillations.
        Such oscillations are called \textit{optomechanical self-oscillations}, which is a characteristic feature in the blue-detuned regime.
        \\

        \textbf{Exercise 7:} What happens when $\delta \omega_{m} < - \omega_{m}$ in the red-detuned regime?
        
        \textbf{Answer:} As the mechanical oscillator is further cooled, the decreasing spring constant may lead to a negative curvature of the intrinsic harmonic potential.
        This might lead to \textit{parametric instability}, which is a characteristic feature in the red-detuned regime.
        \\

        \textbf{Exercise 8:} Derive Equations \eqref{eqn:quantum_equations}.
        
        \textbf{Solution:} Substituting $a = \alpha + \delta a$ and $b = \beta + \delta b$ in Equations \eqref{eqn:oms_qle_eom},
        \begin{subequations}
            \begin{eqnarray}
                \dot{\alpha} + \delta \dot{a} & = & - \left( \frac{\kappa}{2} - i \Delta_{0} \right) \alpha - \left( \frac{\kappa}{2} - i \Delta_{0} \right) \delta a \nonumber \\
                && + i g_{0} \alpha \left( \beta^{*} + \beta \right) + i g_{0} \alpha \left( \delta b^{\dagger} + \delta b \right) \nonumber \\
                && + i g_{0} \delta a \left( \delta b^{\dagger} + \delta b \right) + A_{l} + \sqrt{\kappa} a_{in}, \nonumber \\
                \dot{\beta} + \delta \dot{b} & = & - \left( \frac{\gamma}{2} + i \omega_{m} \right) \beta - \left( \frac{\gamma}{2} + i \omega_{m} \right) \delta b \nonumber \\
                && + i g_{0} \alpha^{*} \alpha + i g_{0} \left( \alpha \delta a^{\dagger} + \alpha^{*} \delta a \right) \nonumber \\
                && + i g_{0} \delta a^{\dagger} \delta a + \sqrt{\gamma} b_{in}. \nonumber
            \end{eqnarray}
        \end{subequations}
        
        Using Equations \eqref{eqn:classical_dynamics} and ignoring the second-order terms in $\delta a$ and $\delta b$, we obtain Equations \eqref{eqn:quantum_equations}.
        \\

        \textbf{Exercise 9:} Obtain the expressions for the drift matrix $\mathbf{A}$ and the noise vector $\mathbf{n}$.

        \textbf{Solution:} Using Equation \eqref{eqn:quantum_equations}, and the fluctuation quadratures, we obtain,
        \begin{subequations}
            \begin{eqnarray}
                \delta \dot{X} & = & - \frac{\kappa}{2} \delta X - \Delta \delta Y - 2 g_{I} \delta Q + \sqrt{\kappa} X_{in}, \nonumber \\
                \delta \dot{Y} & = & \Delta \delta X - \frac{\kappa}{2} \delta Y + 2 g_{R} \delta Q + \sqrt{\kappa} Y_{in}, \nonumber \\
                \delta \dot{Q} & = & - \frac{\gamma}{2} \delta Q + \omega_{m} \delta P + \sqrt{\gamma} Q_{in}, \nonumber \\
                \delta \dot{P} & = & 2 g_{R} \delta X + 2 g_{I} \delta Y - \omega_{m} \delta Q - \frac{\gamma}{2} \delta P \nonumber \\
                && + \sqrt{\gamma} P_{in}, \nonumber
            \end{eqnarray}
        \end{subequations}
        where $g_{R}$ ($g_{I}$) is the real (imaginary) component of $g$ and $X_{in} = (a_{in}^{\dagger} + a_{in}) / \sqrt{2}, Y_{in} = i (a_{in}^{\dagger} - a_{in}) / \sqrt{2}, Q_{in} = (b_{in}^{\dagger} + b_{in}) / \sqrt{2}, P_{in} = i (b_{in}^{\dagger} - b_{in}) / \sqrt{2}$ are the noise quadratures governed by the commutation relations in Equations \eqref{eqn:oms_qle_bcr}.

        Now, using $\mathbf{u} = {(\delta X, \delta Y, \delta Q, \delta P)}^{T}$ as the vector for the quadratures, we can write the above equations in vector form of Equation \eqref{eqn:quantum_linear_compact}, where the drift matrix
        \begin{eqnarray}
            \mathbf{A} & = & \left(
            \begin{array}{cccc}
                - \frac{\kappa}{2} & - \Delta & - 2 g_{I} & 0 \\
                - \Delta & - \frac{\kappa}{2} & 2 g_{R} & 0 \\
                0 & 0 & - \frac{\gamma}{2} & \omega_{m} \\
                2 g_{R} & 2 g_{I} & - \omega_{m} & - \frac{\gamma}{2}
            \end{array}
            \right), \nonumber
        \end{eqnarray}
        and the corresponding vector of the noises is $\mathbf{n} = ( \sqrt{\kappa} X_{in}, \sqrt{\kappa} Y_{in}, \sqrt{\gamma} Q_{in}, \sqrt{\gamma} P_{in} )$.
        \\

        \textbf{Exercise 10:} Derive Equation \eqref{eqn:quantum_linear_int}.

        \textbf{Solution:} Using the unitary transformation $U = \mathrm{exp} [ - i \Delta \delta a^{\dagger} \delta a t + i \omega_{m} \delta b^{\dagger} \delta b t ]$ in Equation \eqref{eqn:quantum_linear_simple} and rearranging the terms, one can obtain Equation \eqref{eqn:quantum_linear_int}.
        \\
                
        \textbf{Exercise 11:} Derive Equation \eqref{eqn:quantum_phenomena_dynamics}.

        \textbf{Solution:} A formal solution for Equation \eqref{eqn:quantum_linear_compact} is
        \begin{equation}
            \mathbf{u}_{i} (t) = \sum_{k} \mathbf{M}_{ik} (t, t_{0}) \mathbf{u}_{k} (t_{0}) + \sum_{k} \int_{t_{0}}^{t} ds \mathbf{M}_{ik} (t, s) \mathbf{n}_{k} (s), \nonumber
        \end{equation}
        where $\dot{\mathbf{M}} (t, t_{0}) = \mathbf{A} (t) \mathbf{M} (t, t_{0})$ with $\mathbf{M} (t, t) = \mathbb{1}$.
        Now, substituting Equation \eqref{eqn:quantum_linear_compact} in Equation \eqref{eqn:quantum_phenomena_v}, we can obtain the dynamics of each correlation matrix element as
        \begin{eqnarray}
            \dot{\mathbf{V}}_{ij} (t) & = & \sum_{k} \mathbf{A}_{ik} (t) \mathbf{V}_{kj} (t) + \sum_{l} \mathbf{A}_{jl} (t) \mathbf{V}_{il} (t) \nonumber \\
            && + \frac{1}{2} \langle \mathbf{n}_{i} (t) \mathbf{u}_{j} (t) + \mathbf{u}_{j} (t) \mathbf{n}_{i} (t) \rangle \nonumber \\
            && + \frac{1}{2} \langle \mathbf{u}_{i} (t) \mathbf{n}_{j} (t) + \mathbf{n}_{j} (t) \mathbf{u}_{i} (t) \rangle, \nonumber
        \end{eqnarray}
        
        Using the fact that the input fluctuations have zero mean, we can rewrite the final expectation term as
        \begin{eqnarray}
            && \sum_{k} \frac{1}{2} \int_{t_{0}}^{t} ds \mathbf{M}_{ik} (t, s) \langle \mathbf{n}_{k} (s) \mathbf{n}_{j} (t) + \mathbf{n}_{j} (t) \mathbf{n}_{k} (s) \rangle \nonumber \\
            && = \sum_{k} \int_{t_{0}}^{t} ds \mathbf{M}_{ik} (t, s) \mathbf{D}_{kj} \delta (s - t) \nonumber \\
            && = \frac{1}{2} \sum_{k} \mathbf{M}_{ik} (t, t) \mathbf{D}_{kj} = \frac{1}{2} \mathbf{D}_{ij} \nonumber
        \end{eqnarray}
        where the noise matrix $\mathbf{D}$ is defined in such a way that $\langle \mathbf{n}_{i} (t) \mathbf{n}_{j} (t^{\prime}) + \mathbf{n}_{j} (t^{\prime}) \mathbf{n}_{i} (t) \rangle / 2 = \mathbf{D}_{ij} \delta (t - t^{\prime})$.

        Similarly, it can be shown that $\langle \mathbf{n}_{i} (t) \mathbf{u}_{j} (t) + \mathbf{u}_{j} (t) \mathbf{n}_{i} (t) \rangle / 2 = \mathbf{D}_{ij} / 2$.
        Thus, we obtain
        \begin{eqnarray}
            \dot{\mathbf{V}}_{ij} (t) = \left[ \mathbf{A} (t) \mathbf{V} (t) \right]_{ij} + \left[ \mathbf{V} (t) \mathbf{A}^{T} (t) \right]_{ij} + \mathbf{D}_{ij} \nonumber
        \end{eqnarray}
        which is analogous to Equation \eqref{eqn:quantum_phenomena_dynamics}.

    \bibliographystyle{ieeetr}
    \bibliography{references}

\begin{thebibliography}{10}

\bibitem{RevModPhys.86.1391}
M.~Aspelmeyer, T.~J. Kippenberg, and F.~Marquardt, ``Cavity optomechanics,''
  {\em Rev. Mod. Phys.}, vol.~86, pp.~1391--1452, 2014.

\bibitem{SBH.CavityOptomechanics.Aspelmeyer}
M.~Aspelmeyer, T.~J. Kippenberg, and F.~Marquardt, {\em Cavity Optomechanics:
  Nano- and Micromechanical Resonators Interacting with Light}.
\newblock Springer Berlin Heidelberg, 2014.

\bibitem{RevSciInstrum.76.061101}
K.~L. Ekinci and M.~L. Roukes, ``Nanoelectromechanical systems,'' {\em Rev.
  Sci. Instrum.}, vol.~76, p.~061101, May 2005.

\bibitem{PhysikZ.10.817}
A.~Einstein, ``On the development of our understanding of the nature and
  composition of radiation,'' {\em Physik. Z.}, vol.~10, pp.~817--825, 1909.

\bibitem{SovPhysJETP.25.653}
V.~B. Braginskii and A.~B. Manukin, ``Pondermotive effects of electromagnetic
  radiation,'' {\em Sov. Phys. JETP}, vol.~25, pp.~653--655, 1967.

\bibitem{NatPhys.18.15}
S.~Barzanjeh, A.~Xuereb, S.~Gr\:{o}blacher, M.~Paternostro, C.~A. Regal, and
  E.~M. Weig, ``Optomechanics for quantum technologies,'' {\em Nat. Phys.},
  vol.~18, pp.~15--24, Dec 2021.

\bibitem{TaylorFrancis.QuantumOptomechanics.Bowen}
W.~P. Bowen and G.~J. Milburn, {\em Quantum Optomechanics}.
\newblock Taylor \& Francis, 2015.

\bibitem{AVSQuantumSci.3.015901}
A.~K. Sarma, S.~Chakraborty, and S.~Kalita, ``Continuous variable quantum
  entanglement in optomechanical systems: A short review,'' {\em AVS Quantum
  Sci.}, vol.~3, p.~015901, 2021.

\bibitem{PhysicsToday.65.29}
M.~Aspelmeyer, P.~Meystre, and K.~Schwab, ``Quantum optomechanics,'' {\em
  Physics Today}, vol.~65, p.~29, 2012.

\bibitem{NatNanotechnol.13.11}
L.~Midolo, A.~Schliesser, and A.~Fiore, ``Nano-opto-electro-mechanical
  systems,'' {\em Nat. Nanotechnol.}, vol.~13, pp.~11--18, 2018.

\bibitem{PhysRevLett.40.729}
A.~Ashkin, ``Trapping of atoms by resonance radiation pressure,'' {\em Phys.
  Rev. Lett.}, vol.~40, pp.~729--732, Mar 1978.

\bibitem{PhysRevA.31.3761}
C.~W. Gardiner and M.~J. Collett, ``Input and output in damped quantum systems:
  Quantum stochastic differential equations and the master equation,'' {\em
  Phys. Rev. A}, vol.~31, pp.~3761--3774, 1985.

\bibitem{PhysRevLett.51.1550}
A.~Dorsel, J.~D. McCullen, P.~Meystre, E.~Vignes, and H.~Walther, ``Optical
  bistability and mirror confinement induced by radiation pressure,'' {\em
  Phys. Rev. Lett.}, vol.~51, pp.~1550--1553, 1983.

\bibitem{PhysRevA.35.5288}
E.~X. DeJesus and C.~Kaufman, ``Routh-hurwitz criterion in the examination of
  eigenvalues of a system of nonlinear ordinary differential equations,'' {\em
  Phys. Rev. A}, vol.~35, pp.~5288--5290, 1987.

\bibitem{NewJPhys.22.013049}
T.~F. Roque, F.~Marquardt, and O.~M. Yevtushenko, ``Nonlinear dynamics of
  weakly dissipative optomechanical systems,'' {\em New J. Phys.}, vol.~22,
  p.~013049, Jan 2020.

\bibitem{PhysRevLett.70.548}
H.~M. Wiseman and G.~J. Milburn, ``Quantum theory of optical feedback via
  homodyne detection,'' {\em Phys. Rev. Lett.}, vol.~70, pp.~548--551, 1993.

\bibitem{PhysRevLett.99.093901}
I.~Wilson-Rae, N.~Nooshi, W.~Zwerger, and T.~J. Kippenberg, ``Theory of ground
  state cooling of a mechanical oscillator using dynamical backaction,'' {\em
  Phys. Rev. Lett.}, vol.~99, p.~093901, Aug 2007.

\bibitem{NatPhys.4.415}
A.~Schliesser, R.~Rivi\`{e}re, G.~Anetsberger, O.~Arcizet, and T.~J.
  Kippenberg, ``Resolved-sideband cooling of a micromechanical oscillator,''
  {\em Nat. Phys.}, vol.~4, pp.~415--419, May 2008.

\bibitem{PhysRevLett.103.213603}
A.~Mari and J.~Eisert, ``Gently modulating optomechanical systems,'' {\em Phys.
  Rev. Lett.}, vol.~103, p.~213603, 2009.

\bibitem{PhysRevA.101.053836}
C.-H. Bai, D.-Y. Wang, S.~Zhang, S.~Liu, and H.-F. Wang, ``Strong mechanical
  squeezing in a standard optomechanical system by pump modulation,'' {\em
  Phys. Rev. A}, vol.~101, p.~053836, 2020.

\bibitem{RepProgPhys.72.076901}
B.~P. Abbott {\em et~al.}, ``{LIGO}: the laser interferometer
  gravitational-wave observatory,'' {\em Reports on Progress in Physics},
  vol.~72, p.~076901, jun 2009.

\bibitem{PhysRevLett.123.231108}
F.~Acernese {\em et~al.}, ``Increasing the astrophysical reach of the advanced
  virgo detector via the application of squeezed vacuum states of light,'' {\em
  Phys. Rev. Lett.}, vol.~123, p.~231108, 2019.

\bibitem{PhysRevLett.98.030405}
D.~Vitali, S.~Gigan, A.~Ferreira, H.~Boehm, P.~Tombesi, A.~Guerreiro,
  V.~Vedral, A.~Zeilinger, and M.~Aspelmeyer, ``Optomechanical entanglement
  between a movable mirror and a cavity field,'' {\em Phys. Rev. Lett.},
  vol.~98, no.~3, p.~030405, 2007.

\bibitem{NewJPhys.14.105010}
Y.-D. Wang and A.~A. Clerk, ``Using dark modes for high-fidelity optomechanical
  quantum state transfer,'' {\em New J. Phys.}, vol.~14, p.~105010, Oct 2012.

\bibitem{PhysRevA.93.062339}
G.~D. de~Moraes~Neto, F.~M. Andrade, V.~Montenegro, and S.~Bose, ``Quantum
  state transfer in optomechanical arrays,'' {\em Phys. Rev. A}, vol.~93,
  p.~062339, Jun 2016.

\bibitem{Nature.588.599}
M.~Mirhosseini, A.~Sipahigil, M.~Kalaee, and O.~Painter, ``Superconducting
  qubit to optical photon transduction,'' {\em Nature}, vol.~588, pp.~599--603,
  Dec 2020.

\bibitem{PhysRevA.92.043845}
S.~Huang, ``Quantum state transfer in cavity electro-optic modulators,'' {\em
  Phys. Rev. A}, vol.~92, p.~043845, Oct 2015.

\bibitem{NatPhys.10.321}
R.~W. Andrews, R.~W. Peterson, T.~P. Purdy, K.~Cicak, R.~W. Simmonds, C.~A.
  Regal, and K.~W. Lehnert, ``Bidirectional and efficient conversion between
  microwave and optical light,'' {\em Nat. Phys.}, vol.~10, pp.~321--326, Mar
  2014.

\bibitem{CUP.QuantumComputationQuantumInformation.Nielsen}
M.~A. Nielsen and I.~L. Chuang, {\em Quantum Computation and Quantum
  Information: 10th Anniversary Edition}.
\newblock Cambridge University Press, 2011.

\bibitem{CUP.QuantumInformationTheory.Wilde}
M.~M. Wilde, {\em Quantum Information Theory}.
\newblock Cambridge University Press, 2013.

\bibitem{PhysRevA.86.042306}
U.~Akram, W.~Munro, K.~Nemoto, and G.~J. Milburn, ``Photon-phonon entanglement
  in coupled optomechanical arrays,'' {\em Phys. Rev. A}, vol.~86, p.~042306,
  2012.

\bibitem{PhysRevA.94.053807}
M.~Wang, X.-Y. L\"u, Y.-D. Wang, J.~Q. You, and Y.~Wu, ``Macroscopic quantum
  entanglement in modulated optomechanics,'' {\em Phys. Rev. A}, vol.~94,
  p.~053807, 2016.

\bibitem{PhysRevA.84.042342}
S.~Barzanjeh, D.~Vitali, P.~Tombesi, and G.~J. Milburn, ``Entangling optical
  and microwave cavity modes by means of a nanomechanical resonator,'' {\em
  Phys. Rev. A}, vol.~84, p.~042342, 2011.

\bibitem{PhysRevLett.114.080503}
S.~Barzanjeh, S.~Guha, C.~Weedbrook, D.~Vitali, J.~H. Shapiro, and
  S.~Pirandola, ``Microwave quantum illumination,'' {\em Phys. Rev. Lett.},
  vol.~114, p.~080503, 2015.

\bibitem{NewJPhys.22.063041}
J.~Li and S.~Gr\"{o}blacher, ``Stationary quantum entanglement between a
  massive mechanical membrane and a low frequency lc circuit,'' {\em New J.
  Phys.}, vol.~22, p.~063041, 2020.

\bibitem{PhysRevA.106.043501}
S.~Kalita, S.~Shah, and A.~K. Sarma, ``Significant optoelectrical entanglement
  and mechanical squeezing in a multimodulated optoelectromechanical system,''
  {\em Phys. Rev. A}, vol.~106, p.~043501, Oct 2022.

\bibitem{PBL.Sync.Strogatz}
S.~H. Strogatz, {\em Sync: The Emerging Science of Spontaneous Order}.
\newblock Penguin Books Limited, 2004.

\bibitem{SIP.QuantumCorrSyncMeasures.Glave}
F.~Galve, G.~Luca~Giorgi, and R.~Zambrini, {\em Quantum Correlations and
  Synchronization Measures}, pp.~393--420.
\newblock Cham: Springer International Publishing, 2017.

\bibitem{PhysRevA.91.012301}
V.~Ameri, M.~Eghbali-Arani, A.~Mari, A.~Farace, F.~Kheirandish, V.~Giovannetti,
  and R.~Fazio, ``Mutual information as an order parameter for quantum
  synchronization,'' {\em Phys. Rev. A}, vol.~91, p.~012301, 2015.

\bibitem{PhysRevA.99.043804}
M.~Koppenh\"{o}fer and A.~Roulet, ``Optimal synchronization deep in the quantum
  regime: Resource and fundamental limit,'' {\em Phys. Rev. A}, vol.~99,
  p.~043804, 2019.

\bibitem{PhysRevLett.111.103605}
A.~Mari, A.~Farace, N.~Didier, V.~Giovannetti, and R.~Fazio, ``Measures of
  quantum synchronization in continuous variable systems,'' {\em Phys. Rev.
  Lett.}, vol.~111, p.~103605, 2013.

\bibitem{PhysRevResearch.2.043287}
N.~Jaseem, M.~Hajdu\v{s}ek, P.~Solanki, L.-C. Kwek, R.~Fazio, and
  S.~Vinjanampathy, ``Generalized measure of quantum synchronization,'' {\em
  Phys. Rev. Research}, vol.~2, p.~043287, 2020.

\bibitem{PhysRevLett.109.233906}
M.~Zhang, G.~S. Wiederhecker, S.~Manipatruni, A.~Barnard, P.~McEuen, and
  M.~Lipson, ``Synchronization of micromechanical oscillators using light,''
  {\em Phys. Rev. Lett.}, vol.~109, p.~233906, 2012.

\bibitem{PhysRevA.99.033818}
C.-G. Liao, R.-X. Chen, H.~Xie, M.-Y. He, and X.-M. Lin, ``Quantum
  synchronization and correlations of two mechanical resonators in a
  dissipative optomechanical system,'' {\em Phys. Rev. A}, vol.~99, p.~033818,
  2019.

\bibitem{SciRep.3.01439}
G.~Manzano, F.~Galve, G.~Giorgi, E.~Hernández-García, and R.~Zambrini,
  ``Synchronization, quantum correlations and entanglement in oscillator
  networks,'' {\em Sci. Rep.}, vol.~3, p.~1439, 2013.

\bibitem{OptExpress.24.012336}
T.~Li, T.-Y. Bao, Y.-L. Zhang, C.-L. Zou, X.-B. Zou, and G.-C. Guo,
  ``Long-distance synchronization of unidirectionally cascaded optomechanical
  systems,'' {\em Opt. Express}, vol.~24, no.~11, pp.~12336--12348, 2016.

\bibitem{CommunNonlinearSciNumerSimulat.42.121}
W.~Li, F.~Zhang, C.~Li, and H.~Song, ``Quantum synchronization in a star-type
  cavity qed network,'' {\em Commun. Nonlinear Sci. Numer. Simulat.}, vol.~42,
  pp.~121--131, 2017.

\bibitem{PhysRevE.95.022204}
W.~Li, C.~Li, and H.~Song, ``Quantum synchronization and quantum state sharing
  in an irregular complex network,'' {\em Phys. Rev. E}, vol.~95, p.~022204,
  2017.

\bibitem{Photonics.4.48}
K.~Sina, ``Higher-order interactions in quantum optomechanics: Analytical
  solution of nonlinearity,'' {\em Photonics}, vol.~4, p.~48, 2017.

\bibitem{SciRep.8.16676}
S.~Khorasani, ``Higher-order interactions in quantum optomechanics: Analysis of
  quadratic terms,'' {\em Sci. Rep.}, vol.~8, p.~16676, 2018.

\bibitem{PhysRevA.82.021806}
A.~Nunnenkamp, K.~B\o{}rkje, J.~G.~E. Harris, and S.~M. Girvin, ``Cooling and
  squeezing via quadratic optomechanical coupling,'' {\em Phys. Rev. A},
  vol.~82, p.~021806, Aug 2010.

\bibitem{NewJPhys.18.013043}
T.~Weiss, A.~Kronwald, and F.~Marquardt, ``Noise-induced transitions in
  optomechanical synchronization,'' {\em New J. Phys.}, vol.~18, p.~013043,
  2016.

\bibitem{PhysRevA.96.043832}
M.~Mikkelsen, T.~Fogarty, J.~Twamley, and T.~Busch, ``Optomechanics with a
  position-modulated kerr-type nonlinear coupling,'' {\em Phys. Rev. A},
  vol.~96, p.~043832, Oct 2017.

\bibitem{PhysRevE.83.056202}
C.~P. Meaney, R.~H. McKenzie, and G.~J. Milburn, ``Quantum entanglement between
  a nonlinear nanomechanical resonator and a microwave field,'' {\em Phys. Rev.
  E}, vol.~83, p.~056202, May 2011.

\bibitem{PhysRevA.88.023853}
J.-Q. Liao and F.~Nori, ``Photon blockade in quadratically coupled
  optomechanical systems,'' {\em Phys. Rev. A}, vol.~88, p.~023853, Aug 2013.

\bibitem{PhysRevA.98.013826}
B.~Sarma and A.~K. Sarma, ``Unconventional photon blockade in three-mode
  optomechanics,'' {\em Phys. Rev. A}, vol.~98, p.~013826, Jul 2018.

\bibitem{PhysRevA.81.041803}
G.~S. Agarwal and S.~Huang, ``Electromagnetically induced transparency in
  mechanical effects of light,'' {\em Phys. Rev. A}, vol.~81, p.~041803, 2010.

\bibitem{PhysRevA.88.013804}
M.~Karuza, C.~Biancofiore, M.~Bawaj, C.~Molinelli, M.~Galassi, R.~Natali,
  P.~Tombesi, G.~Di~Giuseppe, and D.~Vitali, ``Optomechanically induced
  transparency in a membrane-in-the-middle setup at room temperature,'' {\em
  Phys. Rev. A}, vol.~88, p.~013804, Jul 2013.

\bibitem{PhysRevA.101.043820}
X.-B. Yan, ``Optomechanically induced transparency and gain,'' {\em Phys. Rev.
  A}, vol.~101, p.~043820, Apr 2020.

\bibitem{PhysRevLett.107.043603}
G.~Heinrich, M.~Ludwig, J.~Qian, B.~Kubala, and F.~Marquardt, ``Collective
  dynamics in optomechanical arrays,'' {\em Phys. Rev. Lett.}, vol.~107,
  p.~043603, 2011.

\bibitem{Nature.600.75}
J.~Zhang, B.~Peng, S.~Kim, F.~Monifi, X.~Jiang, Y.~Li, P.~Yu, L.~Liu, Y.-x.
  Liu, A.~Al\'{u}, and L.~Yang, ``Optomechanical dissipative solitons,'' {\em
  Nature}, vol.~600, pp.~75--80, 2021.

\bibitem{RevModPhys.93.025005}
A.~Blais, A.~L. Grimsmo, S.~M. Girvin, and A.~Wallraff, ``Circuit quantum
  electrodynamics,'' {\em Rev. Mod. Phys.}, vol.~93, p.~025005, May 2021.

\bibitem{PhysRevLett.127.113601}
P.~Kumar, T.~Biswas, K.~Feliz, R.~Kanamoto, M.-S. Chang, A.~K. Jha, and
  M.~Bhattacharya, ``Cavity optomechanical sensing and manipulation of an
  atomic persistent current,'' {\em Phys. Rev. Lett.}, vol.~127, p.~113601, Sep
  2021.

\bibitem{FrontPhys.17.42201}
K.~Wang, Y.-P. Gao, R.~Jiao, and W.~Chuan, ``Recent progress on optomagnetic
  coupling and optical manipulation based on cavity-optomagnonics,'' {\em
  Front. Phys.}, vol.~17, p.~42201, 2022.

\end{thebibliography}
	
\end{document}